\documentclass[]{article}
\usepackage{lineno}
\usepackage{amsmath}
\usepackage{amsfonts}
\usepackage{amssymb}
\usepackage{stfloats}
\usepackage{graphicx}
\usepackage{hyperref}
\usepackage{siunitx}
\usepackage{graphicx}
\usepackage{amsmath}
\usepackage{color}
\usepackage{bm}
\usepackage{mathrsfs}
\usepackage{amsbsy}
\usepackage{caption}
\usepackage {ulem}
\usepackage[thicklines]{cancel}
\usepackage{diagbox}
\usepackage{subfigure}

\newcommand{\be}{\begin{equation}}
\newcommand{\ee}{\end{equation}}
\newcommand{\bea}{\begin{eqnarray}}
\newcommand{\eea}{\end{eqnarray}}

\usepackage{chngcntr}
\usepackage{cite}
\usepackage[a4paper, top=1.9cm, bottom=4.3cm, left=2.5cm, right=2.5cm]{geometry}

\usepackage{hyperref}
\usepackage[T1]{fontenc}
\usepackage{float} 
\usepackage[utf8]{inputenc}
\usepackage{caption}
\usepackage{graphics}
\usepackage{wrapfig}
\usepackage{url}        
\usepackage{bm}

\usepackage{etoolbox}
\makeatletter
\patchcmd{\thebibliography}
  {\advance\leftmargin\labelsep}
  {\setlength{\itemsep}{0.5ex}  
   \setlength{\parsep}{0pt}
   \advance\leftmargin\labelsep}
  {}{}
\makeatother

\usepackage{subcaption}
\usepackage{breqn}
\usepackage{authblk}

\begin{document}

\title{Quasi-Normal Modes and Nonlinear Electrodynamics in Black Hole Phase Transitions}
\author[1]{Zi-Yu Hou}
\author[1,2]{Yu-Qi Lei\thanks{Corresponding author: yuqi_lei@shu.edu.cn}}
\author[1]{Xian-Hui Ge\thanks{Corresponding author:  gexh@shu.edu.cn}}
\affil[1]{Department of Physics, College of Sciences, Shanghai University, 99 Shangda Road, 200444 Shanghai, China}
\affil[2]{Department of Mathematics, College of Sciences, Shanghai University, 99 Shanghai Road, 200444 Shanghai, China}
\date{}
\maketitle

\begin{abstract}
We investigate the connection between thermodynamic phase transitions and quasi-normal modes (QNMs) in charged black holes with a positive curvature constant, within the framework of $F(R)$-Euler-Heisenberg gravity. Nonlinear electromagnetic fields lead to rich thermodynamic phase structures and significantly affect the QNMs of massless scalar fields. By analyzing the QNMs spectrum, we find that the transition point marking the disappearance of the divergence in the QNMs slope parameter $K$ aligns with the change of the thermodynamic phase structure described by the heat capacity, within the bounds of computational uncertainty. This precise matching holds under variations of the curvature parameter and charge. Furthermore, we show that larger angular quantum number $l$ diminishes this correspondence, while higher overtone number $n$ restores it beyond a threshold. These findings demonstrate that thermodynamic phase transitions of black holes carry embedded dynamical information, uncovering a fundamental link between black hole thermodynamic and dynamical properties.
\end{abstract}

\section{Introduction}\label{sec1}

The thermodynamic properties of black holes reflect the quantum nature of the black hole horizons. The relation between the horizon area and the black hole entropy \cite{PhysRevD.7.2333,Bardeen1973}, the effective temperature revealed by Hawking radiation \cite{HAWKING1974,Hawking1975}, and other related findings \cite{PhysRevD.15.2752,PhysRevD.48.R3427,PhysRevLett.132.141501,PhysRevLett.82.2828} all suggest that black holes are not merely classical geometric structures, but physical objects that may contain a large number of quantum degrees of freedom. With these insights, black hole thermodynamics becomes part of a broader theoretical framework \cite{PhysRevLett.75.1260,PADMANABHAN200549,Padmanabhan_2010}, serving as an important basis for exploring the quantum nature of spacetime. After these basic frameworks of black hole thermodynamics, studies on the phase transition of black holes rapidly developed. In an asymptotically flat background, black holes exhibit thermodynamic phase transitions and rich critical behaviors \cite{Davies:1977bgr,Davies:1978zz,Lu:2010xt}. In the presence of a negative cosmological constant, namely the Anti-de Sitter (AdS) spacetime, charged black holes demonstrate Van der Waals-like phase transition structures \cite{Kubizňák2012,Cai2013,Xu2014,Zou:2014mha,Lan2015,Miao:2018fke,Xu:2020ubo,Xu:2021qyw,Xu:2022jyp,PhysRevD.85.024017,PhysRevD.87.044014,PhysRevD.110.046013,PhysRevD.111.064006}, which can reveal physical insights related to quantum chaos and critical phenomena by analogy with strongly correlated systems \cite{Cao1,Cao2,Cao3,Cao4}, and when connected to holography and transport can further refine phase structure and stability conditions in the presence of higher derivative corrections \cite{Ge:2008ni,Ge:2009eh,Shu:2008yd,Ge:2014aza,Tian:2014goa,Chen:2022tfy,Chen:2023pgs}. In de Sitter (dS) spacetime, the presence of black hole horizons and cosmological horizons renders the thermodynamic properties more complicated. By considering the combined contributions of all horizons, a global thermodynamic framework can be constructed \cite{Urano_2009,Tsallis2013}; alternatively, by focusing solely on the black hole event horizon, one can also analyze its local thermodynamic stability and phase transition behavior \cite{PhysRevD.73.084009,PhysRevD.78.124012}. 
Connections between gravitational theories in dS spacetime and the dynamics of unstable D-branes provide geometric perspectives on phase transitions and critical behavior in dS backgrounds \cite{Li:2015qmf}.
The rich thermodynamic structures of black holes in various backgrounds have attracted widespread attention. However, the microscopic origin of these properties and the deep connection between black hole geometry and quantum features remain major open questions. 

Quasi-normal modes (QNMs) are regarded as the characteristic ``sound'' of black holes under linear perturbations \cite{Hans-PeterNollert_1999}. With the spectrum encoding the fundamental physical parameters of the spacetime and solely determined by the background geometry, they are often referred to as the ``fingerprints'' of black holes and provide a powerful tool for identifying specific spacetime structures. 
{For more features of QNMs, see the comprehensive review \cite{Konoplya:2011qq}. }
In gravitational wave observations, the ringdown phase dominated by QNMs provides crucial insights into identifying the type and parameters of black holes, and even for testing the validity of general relativity \cite{PhysRevLett.116.061102,Berti_2009,Kokkotas1999,Cunha2018}. From the perspective of classical geometry, the high-frequency limit of the QNMs spectrum has been shown to be closely related to the dynamics of unstable photon orbits, revealing a profound link between wave behavior and geometric optics for both static, spherically symmetric spacetimes \cite{PhysRevD.79.064016,PhysRevLett.81.4293,PhysRevD.86.104006} and axisymmetric spacetimes\cite{Jusufi:2020dhz,Yang:2021zqy,Pedrotti:2024znu}. Furthermore, in AdS spacetimes, QNMs play an important role in holographic duality, where they correspond to the poles of retarded Green functions and encode information about thermalization rates and quantum chaos \cite{PhysRevD.62.024027,PhysRevD.64.064024,PhysRevLett.98.091601}. 
As a dynamical characteristic of black holes, the study of QNMs can provide a wealth of information about black holes. 
 
In recent years, it has been suggested that QNMs may encode information about the thermodynamic properties of black holes \cite{Jing:2008an,PhysRevD.77.087501,Liu2014,PhysRevD.97.026014,Mahapatra2016,Zhao2025,2025arXiv250404995Z,LAN2021115539,GUO2024138801}. The investigations began with four-dimensional Reissner-Nordström (RN) black holes, in which it was found that the first maximum of the real part of the QNMs frequencies coincides with the thermodynamic critical point \cite{Jing:2008an}. Although this correspondence was later shown to be not universal, failing for Kerr black holes, Schwarzschild-AdS (SAdS) black holes, and other types of perturbations in RN backgrounds \cite{PhysRevD.77.087501}, the idea that QNMs encode thermodynamic features has persisted and been further developed. In RN-AdS backgrounds, a dramatic change in the slope of QNMs frequencies is observed near the critical point of the Van der Waals-like phase transition \cite{Liu2014}. With the inclusion of a dilaton field or Weyl correction, the QNMs frequencies also capture the essence of black hole phase transitions \cite{PhysRevD.97.026014,Mahapatra2016}. In studies of supercritical phase transition, QNMs exhibit correlations with the crossover lines \cite{Zhao2025,2025arXiv250404995Z}. In various regular black hole models whose QNMs show distinct features from those of singular black holes, the consistency still holds \cite{LAN2021115539,GUO2024138801}. {In critical gravity, black hole quasinormal modes may exhibit sensitivity to structural parameters that also affect the greybody factor \cite{Lin:2024ubg}.} All of these results provide further support for the idea that QNMs may encode thermodynamic information about black holes. The consistency between QNMs and black hole thermodynamics merits further investigation. 

In this work, we focus on the $F(R)$-Euler-Heisenberg black holes, from a completely new perspective, explore the possible connection between their thermodynamic and dynamical properties. Compared with previous studies, we focus on the consistency between the transitions in the QNMs behavior and the variations of the thermodynamic phase structure, rather than solely on the phase transition points. The presence of $F(R)$ corrections and nonlinear electromagnetic fields of Euler-Heisenberg theory enriches the thermodynamic phase structure of the system \cite{SEKHMANI2024101701}. After computing the QNMs, analyzing their behavior and relating the results to the thermodynamic phase structure of the black hole, we find a relationship between them. The transition point of QNMs described by the slope parameter coincides with the thermodynamic phase structure in the black hole heat capacity, and this correspondence remains under variations of other black hole parameters. In a word, different QNMs slope behavior may correspond different thermodynamic phase structures. In addition, an increase in the angular quantum number $l$ tends to suppress this correspondence, whereas a higher overtone number $n$ helps to restore it. These results show that the dynamical features of black holes are deeply imprinted in their thermodynamic phase transitions, offering a novel and perspective on the interplay between black hole dynamics and thermodynamics.

The remainder of this paper is organized as follows. In Section \ref{sec2}, we introduce the geometry of the $F(R)$-Euler-Heisenberg black hole and discuss its basic thermodynamic properties. Section \ref{sec3} is devoted to the computation and analysis of QNMs. In Section \ref{sec4}, we explore the consistency between the thermodynamic phase structure and the behavior of QNMs, with particular attention to the influence of varying black hole parameters. Finally, in Section \ref{sec5}, we present our conclusions. Appendix \ref{appendix1} offers a brief examination of the blackening factor, illustrating its structural features, Appendix \ref{appendix2} presents the detailed derivation of the QNMs computation using pseudo spectral method, and Appendix \ref{appendix3} contains further discussions on the effective potential and the Lyapunov exponent associated with the black hole photon sphere.

\section{Black hole geometry and its thermodynamics}\label{sec2}

In this section, we review the geometric structure and thermodynamic properties of $F(R)$-Euler-Heisenberg black holes. These solutions arise from nonlinear electrodynamics and modified gravity theories, and they exhibit a richer horizon structure and thermodynamic behavior. {The gravitational properties of black holes with nonlinear electromagnetic fields or $F(R)$ gravity corrections have been investigated, including their photon spheres, shadow formation, and evaporation processes \cite{AraujoFilho:2024xhm,AraujoFilho:2024lsi,AraujoFilho:2025hnf}.} Using the natural unit system $G=c=\hbar=1$, the action in the framework of $F(R)$-Euler-Heisenberg theory takes the following form \cite{SEKHMANI2024101701}:
\begin{equation}
    \mathcal{I}_{F(R)}=\frac{1}{16\pi}\int_{\mathcal{M}}\mathrm{d}^4 x \sqrt{-g}\Big(F(R)-\mathcal{L}(X,Y)\Big),
\end{equation}
where $F(R)=R+\tilde{f}(R)$ is the rescaled gravitational function in terms of the Ricci scalar $R$, and $\mathcal{L}(X,Y)$ is the Lagrangian of the non-linear electrodynamic theory, which depends on the only two independent relativistic invariants constructed with the Faraday tensor for the Maxwell field in four dimensions. It takes the form 
\begin{equation}
    \mathcal{L}(X,Y)=-X+\frac{\lambda}{2}X^2+\frac{7\lambda}{8}Y^2,
    \label{eq:lagrangian}
\end{equation}
in which $\lambda$ is the Euler-Heisenberg parameter, $X$ and $Y$ are the independent scalar and pseudo-scalar,  expressed as
\begin{equation}
\begin{aligned}
    &X=\frac{1}{4}F_{\mu\nu}F^{\mu\nu}=\frac{1}{2}\left(\mathbf{B}^2-\mathbf{E}^2\right),\\ 
    &Y=\frac{1}{4}F_{\mu\nu}\,^*F^{\mu\nu}=\mathbf{E}\cdot\mathbf{B},
\end{aligned}
\end{equation}
where $\mathbf{E}$ and $\mathbf{B}$ are the electric field strength and the magnetic field strength, respectively, with $F_{\mu\nu}$ being the Faraday electromagnetic tensor and $^*F^{\mu\nu}$ its dual, which is defined by
\begin{equation}
\begin{aligned}
    ^*F_{\mu\nu}=\frac{1}{2}\sqrt{-g}\ \epsilon_{\mu\nu\rho\sigma}F^{\rho\sigma},\qquad&\epsilon_{0123}=-1,\\
    ^*F^{\mu\nu}=\frac{1}{2}\frac{1}{\sqrt{-g}}\epsilon^{\mu\nu\rho\sigma}F_{\rho\sigma},\qquad&\epsilon^{0123}=1,
\end{aligned}
\end{equation}
where $\epsilon_{\mu\nu\rho\sigma}$ is completely antisymmetric and satisfies $\epsilon_{\mu\nu\rho\sigma}\epsilon^{\mu\nu\rho\sigma}=-4!$. The corresponding Einstein equations are 
\begin{equation}
    8\pi T_{\mu\nu}=R_{\mu\nu}(1+\tilde{f}_R)-\frac{g_{\mu\nu}h(r)}{2}+(g_{\mu\nu}\nabla^2-\nabla_\mu\nabla_\nu)\tilde{f}_R,
    \label{eq.field}
\end{equation}
where $R_{\mu\nu}$ is the Ricci curvature tensor, $\tilde{f}_R=\frac{\mathrm{d}\tilde{f}(R)}{\mathrm{d}R}$. 
The energy momentum tensor $T_{\mu\nu}$ is given by
\begin{equation}
    T_{\mu\nu}=\frac{1}{4\pi}\bigg(
    g_{\mu\nu}\mathcal{L}-\left(\mathcal{L}_XF_{\mu\sigma}+\mathcal{L}_Y\,^*F_{\mu\sigma}\right)F_\nu^\sigma\bigg),
    \label{eq:emt1}
\end{equation}
while the subscript on $\mathcal{L}$ stands for derivative with respect to the corresponding invariant. The variation with respect to the electromagnetic four-potential $A_\mu$ yields the electromagnetic field equations,
\begin{equation}
    \nabla_\mu\left(\mathcal{L}_XF^{\mu\nu}+\mathcal{L}_Y\,^*F^{\mu\nu}\right)=0.
    \label{eq: max}
\end{equation}
One can introduce a Legendre dual description using the antisymmetric tensor $P_{\mu\nu}$ defined by
\begin{equation}
    \mathrm{d}\mathcal{L}(X,Y)=-\frac{1}{2}P^{\mu\nu}\mathrm{d}F_{\mu\nu}.
\end{equation}
It reads as
\begin{equation}
    P^{\mu\nu}=2\frac{\partial \mathcal{L}}{\partial F_{\mu\nu}}=-\left(\mathcal{L}_XF^{\mu\nu}+\mathcal{L}_Y\,^*F^{\mu\nu}\right),
\end{equation}
and the field equation Eq.(\ref{eq: max}) can be written as
\begin{equation}
    \nabla_\mu P^{\mu\nu}=0.
\end{equation}
The two invariants of the tensor $P_{\mu\nu}$ are
\begin{equation}
    P=-\frac{1}{4}P_{\mu\nu}P^{\mu\nu},\,O=-\frac{1}{4}P_{\mu\nu}\ ^*P^{\mu\nu},
\end{equation}
where $^*P^{\mu\nu}=\frac{1}{2}\sqrt{-g}\epsilon^{\mu\nu\rho\sigma}P_{\rho\sigma}$. The structural function $\mathcal{H}$ is 
\begin{equation}
    \mathcal{H}(P,O)=-\frac{1}{2}P^{\mu\nu}F_{\mu\nu}-\mathcal{L}.
\end{equation}
Irrespective of the second-order and higher-order modes in $\lambda$, it can be expressed as 
\begin{equation}
    \mathcal{H}(P,O)=P-\frac{\lambda}{2}P^2-\frac{7\lambda}{8}O^2.
\end{equation}
The energy-momentum tensor Eq.(\ref{eq:emt1}) can be written in the $P$ frame \cite{SEKHMANI2024101701,ehbh},
\begin{equation}
\begin{aligned}
     T_{\mu\nu}&=
    \frac{1}{4\pi}\bigg(\mathcal{H}_PP_\mu^\beta P_{\nu\beta}+g_{\mu\nu}(2P\mathcal{H}_P+O\mathcal{H}_O-\mathcal{H})\bigg)\\
    &=\frac{1}{4\pi}\left((1-\lambda P)P_\mu^\beta P_{\nu\beta}+g_{\mu\nu}(P-\frac{3}{2}\lambda P^2-\frac{7\lambda}{8}O^2)\right),
\end{aligned}
\end{equation}
while the subscript on $\mathcal{H}$ stands for derivative with respect to the corresponding invariant. The linear Maxwell electrodynamics is recovered when $\lambda = 0$.

The line element of the black hole takes a static, spherical form,
\begin{equation}
    \mathrm{d}s^2 = -h(r)\mathrm{d} t^2+\frac{\mathrm{d}r^2}{h(r)}+r^2(\mathrm{d}\theta^2+\mathrm{sin}^2\theta \mathrm{d}\phi^2),
    \label{eq:met}
\end{equation} 

{The theory requires the adoption of the traceless energy-momentum tensor condition \cite{PhysRevD.80.124011,Moon:2011hq}. Assuming the scalar curvature to be constant, $R = R_0 = \text{const.}$ \cite{Cognola_2005}, the trace of Eq.(\ref{eq.field}) can be written as}
\begin{equation}
R_0\left(1+f_{R_0}\right)-2\left(R_0+\tilde{f}(R_0)\right)=0,
\end{equation}
{in which $\left. f_{R_0}=\frac{\mathrm{d}\tilde{f}(R)}{\mathrm{d}R}\right|_{R=R_0}$. The $F(R)$ term of the theory can be characterized by the two parameters $R_0$ and $f_{R_0}$. The field equations Eq.(\ref{eq.field}) can be reduced to}
\begin{equation}
R_{\mu\nu}(1+f_{R_0})-\frac{g_{\mu\nu}}{4}R_0(1+f_{R_0})=8\pi T_{\mu\nu}.
\end{equation}

{For the nonlinear electromagnetic field, a suitable choice of the electromagnetic tensor can be expressed as}
\begin{equation}
P_{\mu\nu}=\frac{q}{r^2}\left(\delta^0_\mu \delta^1_\nu - \delta^0_\nu \delta^1_\mu\right), 
\label{eq:main ele}
\end{equation}
and the electromagnetic invariants $P$ and $O$ are fixed in
\begin{equation}
P=\frac{q^2}{2r^4},\ O=0,
\end{equation}
{in which the parameter $q$ is a parameter that matches the total charge of the model. 
Based on the metric Eq.(\ref{eq:met}) and the electric main tensor Eq.(\ref{eq:main ele}), the corresponding field equations can be derived as follows}
\begin{equation}
\begin{aligned}
&Eq_{tt}=Eq_{rr}=rh^{\prime}(r)+h(r)+\frac{r^2R_0}{4}-1-\frac{1}{1+f_{R_0}}\left(\frac{\lambda q^2}{4r^6}+\frac{q^2}{r^2}\right),\\
&Eq_{\theta\theta}=Eq_{\phi\phi}=rh^{\prime\prime}(r)+2h’(r)+\frac{rR_0}{2}+\frac{1}{1+f_{R_0}}\left(\frac{3}{4}\frac{\lambda q^4}{r^7}+\frac{q^2}{r^3}\right),
\end{aligned}
\end{equation}
where $Eq_{tt}$, $Eq_{rr}$, $Eq_{\theta\theta}$ and $Eq_{\phi\phi}$ denote the four directional components ($tt$, $rr$, $\theta\theta$, and $\phi\phi$) of the field equations  Eq.(\ref{eq.field}), respectively.

The explicit form of the blackening factor $h(r)$ is finally given by \cite{SEKHMANI2024101701}
\begin{equation}
    h(r)=1-\frac{m_0}{r}-\frac{R_0 r^2}{12}+\frac{1}{1+f_{R_0}}\left( \frac{q^2}{r^2}-\frac{\lambda q^4}{20 r^6} \right),
    \label{eq:metric func}
\end{equation}
in which $R_0$ is a constant scalar curvature $R=R_0=\mathrm{constant}$, $m_0$ stands for the black hole's geometric mass, $\left. f_{R_0}=\frac{\mathrm{d}\tilde{f}(R)}{\mathrm{d}R}\right|_{R=R_0}$ , $q$ is a parameter that matches the total charge of the model and $\lambda$ is the Euler-Heisenberg parameter. By imposing the gauge condition $f_{R_0} = 0$ together with $R_0 = 4\Lambda$, the modified gravity solution naturally reduces to the standard black hole solution in General Relativity with a cosmological constant $\Lambda$. The sign of $R_0$, which is directly linked to the sign of the cosmological constant $\Lambda$, determines the underlying spacetime structure,  specifically, whether the background geometry is asymptotically dS ($R_0 > 0$) or AdS ($R_0 < 0$). Setting the Euler-Heisenberg parameter $\lambda = 0$, the nonlinear electromagnetic correction would vanish, and the system recovers the RN black hole solution with a cosmological constant with the gauge condition $f_{R_0} = 0$ and $R_0 = 4\Lambda$.
\textbf{Throughout the rest of this paper, we focus on the case $R_0 > 0$, which corresponds to a positive cosmological constant.}

\begin{figure}[htp!]
    \centering
    \includegraphics[width=0.5\linewidth]{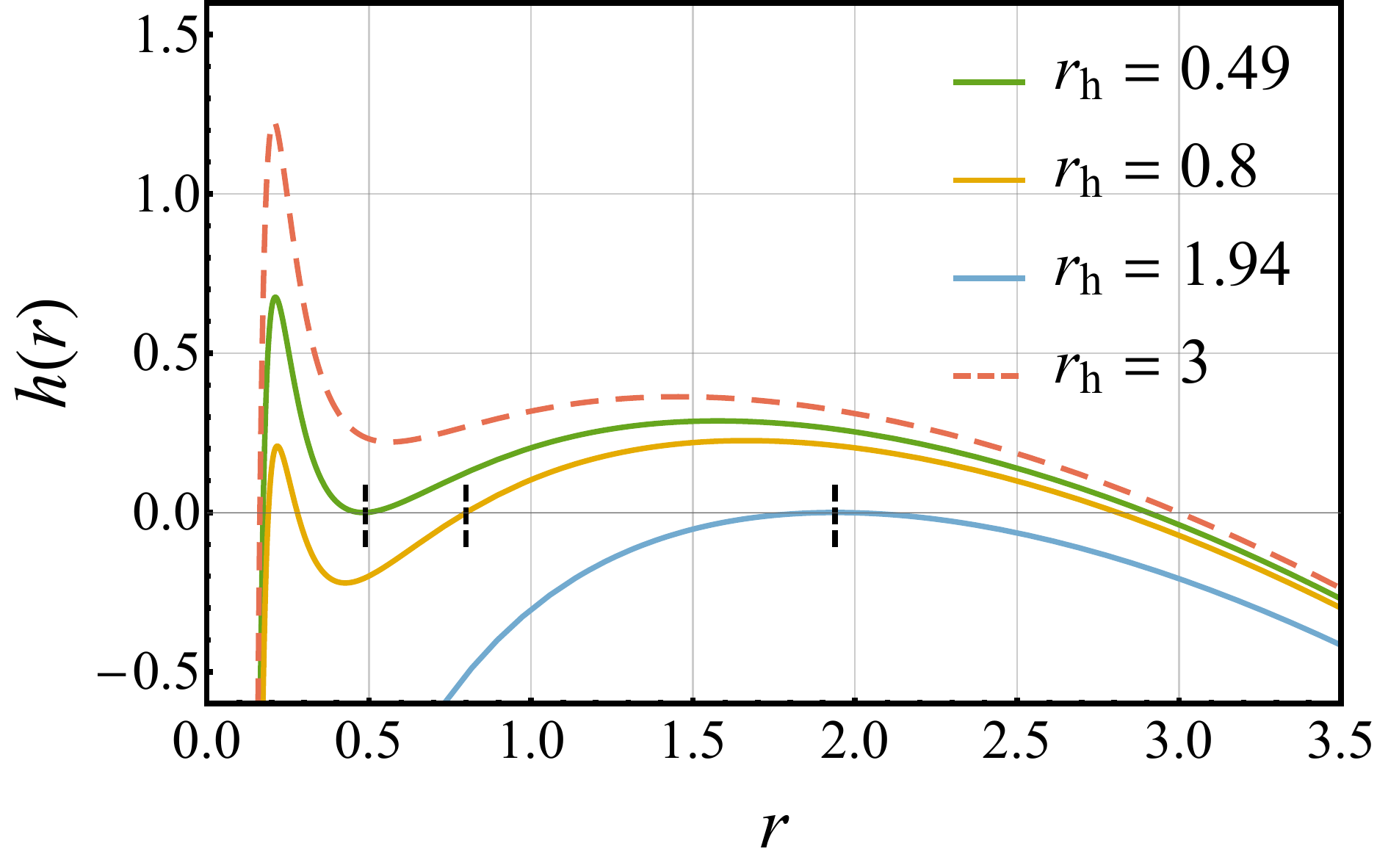}
    \caption{The blackening factor $h(r)$ of $F(R)$-Euler-Heisenberg black holes with $R_0=1$, $q=0.5$, $\lambda=0.03$ and $f_{R_0}=0.1$ for different $r_{\text{h}}$. The red dashed line does not correspond to an actual black hole solution, while the black dashed line marks the event horizon radius $r_{\text{h}}$ of the black hole solution.}
    \label{fig:horizons}
\end{figure}

Through the imposition of $h(r)=0$ at the black hole event horizon $r=r_{\text{h}}$, the mass parameter $m_0$ can be expressed as
\begin{equation}
    m_0=r_{\text{h}}-\dfrac{1}{12}R_0r_{\text{h}}^3+\dfrac{q^2}{\left(1+f_{R_0}\right)r_{\text{h}}}\left(1-\dfrac{\lambda q^2}{20r^4_{\text{h}}}\right).
    \label{eq:mass perfrom}
\end{equation}
The blackening factor structure for $R_0>0$ is shown in Fig. \ref{fig:horizons}, where up to four distinct horizons may exist (the yellow curve). From large to small, the first one is the cosmological horizon $r_{\text{c}}$, followed by the black hole event horizon $r_{\text{h}}$ (the black dashed line), and two inner horizons, $r_{\text{i}}$ and $r_{\text{i}^{\prime}}$, respectively.
More discussions about the effect of the parameters on the black hole geometry are shown in the Appendix \ref{appendix1}.
As shown in Fig. \ref{fig:horizons}, the black hole exhibits two distinct limits. One of them is the extremal black hole limit corresponding to the case of the green line, where the inner horizon coincides with the black hole horizon, $r_{\text{h}}=r_{\text{i}}=r_{\text{extreme}}$. The other one is the Nariai limit corresponding to the case of the blue line where the cosmological horizon coincides with the black hole horizon, $r_{\text{h}}=r_{\text{c}}=r_{\text{Nariai}}$ \cite{Nariai1999}. Beyond these limits, the solution no longer represents a black hole, as indicated by the red dashed line.

\begin{figure}[htp!]
    \centering
    \includegraphics[width=0.5\linewidth]{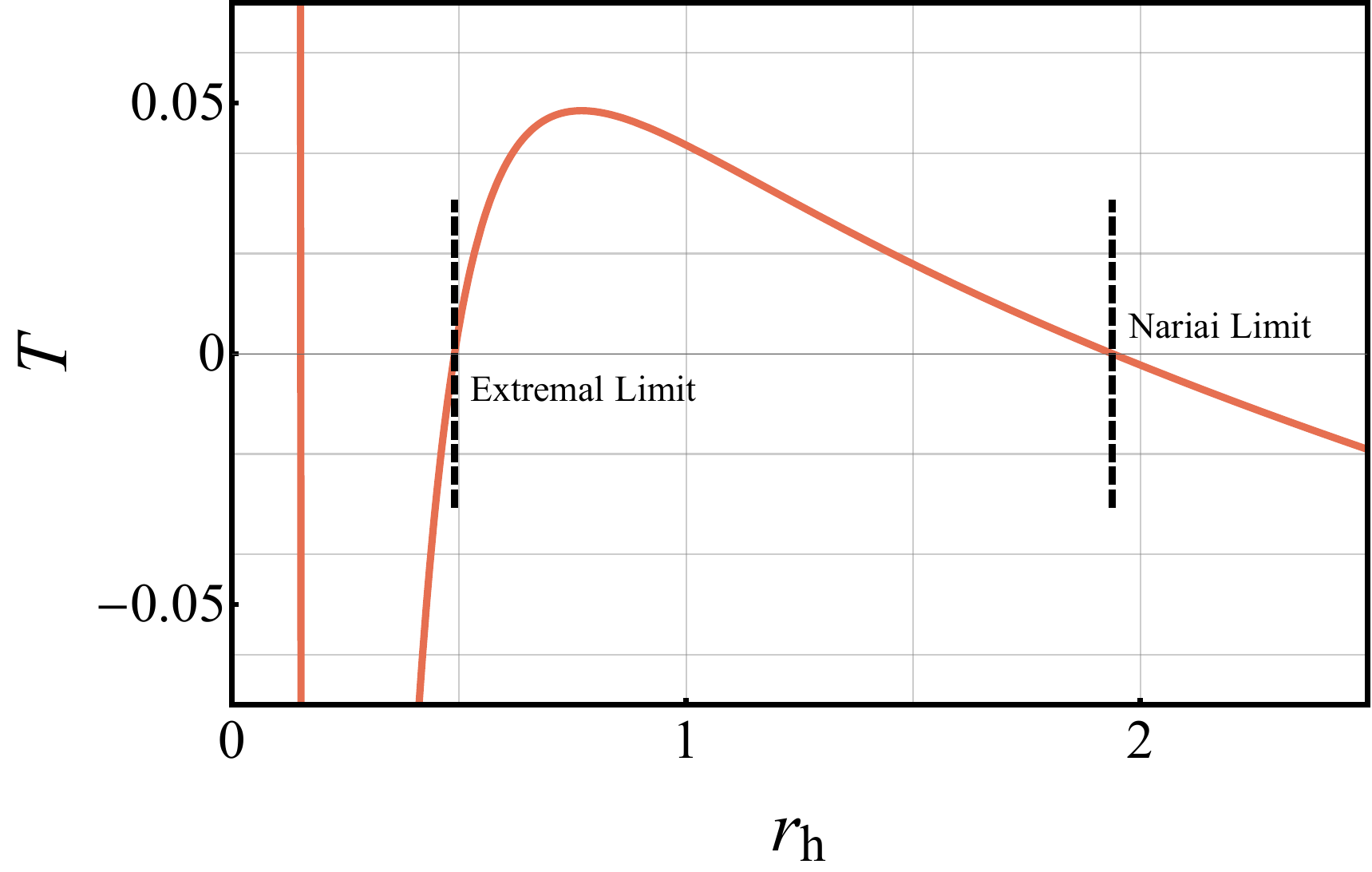}
    \caption{The Hawking temperature $T$ as a function of the black hole event horizon $r_{\text{h}}$ with $R_0=1$, $q=0.5$, $\lambda=0.03$ and $f_{R_0}=0.1$. The two black dashed lines represent the Nariai limit and the extremal limit, respectively, where the Hawking temperature drops to zero $T = 0$. }
    \label{fig:temp}
\end{figure}

We restrict our attention to the thermodynamic behavior of the event horizon to analyze the local properties of the black hole. The Hawking temperature can be represented as
\begin{equation}
    T=\frac{1}{4\pi}\left. \frac{\mathrm{d}h(r)}{\mathrm{d}r}\right|_{r=r_{\text{h}}}=\frac{1}{4\pi r_{\text{h}}}-\frac{R_0r_{\text{h}}}{16\pi}+\frac{q^2}{16\pi\left(1+f_{R_0}\right)r_{\text{h}}^3}\left(\frac{\lambda q^2}{r_{\text{h}}^4}-4\right).
\end{equation}
The specific behavior of the Hawking temperature is illustrated in Fig. \ref{fig:temp}, showing that it remains positive between the two limits, corresponding to physically realizable black holes. Outside the limits, the black hole solutions do not possess physical significance.

It is useful to define a \textbf{rescaled radius} $\tilde{r}_{\text{h}}$ of the black hole event horizon to facilitate subsequent calculations and analyses,
\begin{equation}
    \tilde{r}_{\text{h}}=\frac{r_{\text{h}}-r_{\text{extreme}}}{r_{\text{Nariai}}-r_{\text{extreme}}},
\end{equation}
in which $r_{\text{extreme}}$ stands for the extremal black hole limit, and $r_{\text{Nariai}}$ stands for the Nariai limit. In this way, the black hole behaviors can be described within the interval $\tilde{r}_{\text{h}} \in [0,1]$, while discarding the solutions that lack physical relevance.

For the other thermodynamic quantities, the associated entropy of black holes is
\begin{equation}
    S=\pi(1+f_{R_0})r_{\text{h}}^2.
\end{equation}
The physical charge  $Q$ and the electromagnetic potential $\Phi$ are given as follows,
\begin{equation}
    Q=q\ ,\ \Phi=\int_{r_{\text{h}}}^\infty\mathrm{d}r\ (-F_{tr})=\int_{r_{\text{h}}}^\infty\mathrm{d}r\ P_{tr}\frac{\partial\mathcal{H}}{\partial  P}=\frac{q}{r_{\text{h}}}\left(1-\frac{\lambda q^2}{10r_{\text{h}}^4}\right). 
\end{equation}
By means of the Ashtekar-Magnon-Das (AMD) approach \cite{AbhayAshtekar_2000,Dhurandhar_2005}, the total mass of the black hole can also be obtained as
\begin{equation}
    M=\frac{m_0(1+f_{R_0})}{2}=-\frac{1}{24}\left(1+f_{R_0}\right)r_{\text{h}}(R_0r_{\text{h}}^2-12)-\frac{\lambda q^4-20q^2r_{\text{h}}^4}{40r_{\text{h}}^5}.
\end{equation}
These results are consistent with the first law of black hole thermodynamics,
\begin{equation}
    \mathrm{d}M = T\mathrm{d}S + \Phi\mathrm{d}Q. 
\end{equation}
The heat capacity is given in the form
\begin{equation}
    C_Q=T\left(\frac{\partial S}{\partial T}\right)_Q=\frac{2S^3\big(4\pi^2(1+f_{R_0})Q^2-4\pi(1+f_{R_0})S+R_0S^2\big)-2\pi^4\lambda(1+f_{R_0})^3Q^4S}{7\pi^4\lambda(1+f_{R_0})^3Q^4+S^2\big(-12\pi^2(1+f_{R_0})Q^2+4\pi(1+f_{R_0})S+R_0S^2\big)},
\end{equation}
and the Helmholtz free energy is expressed as
\begin{equation}
    F=M-TS=\frac{\sqrt{S}\sqrt{1+f_{R_0}}}{4\sqrt{\pi}}+\frac{R_0S^{3/2}}{48\pi^{3/2}\sqrt{1+f_{R_0}}}+\frac{\sqrt{\pi}Q^2\sqrt{1+f_{R_0}}\left(60S^2-7\pi^2\lambda Q^2(1+f_{R_0})^2\right)}{80S^{5/2}}
\end{equation}

\begin{figure}[!ht]
\centering

\subfigure{\label{CQ1}
\begin{minipage}[t]{0.3\linewidth}
\centering
\includegraphics[width=1\textwidth]{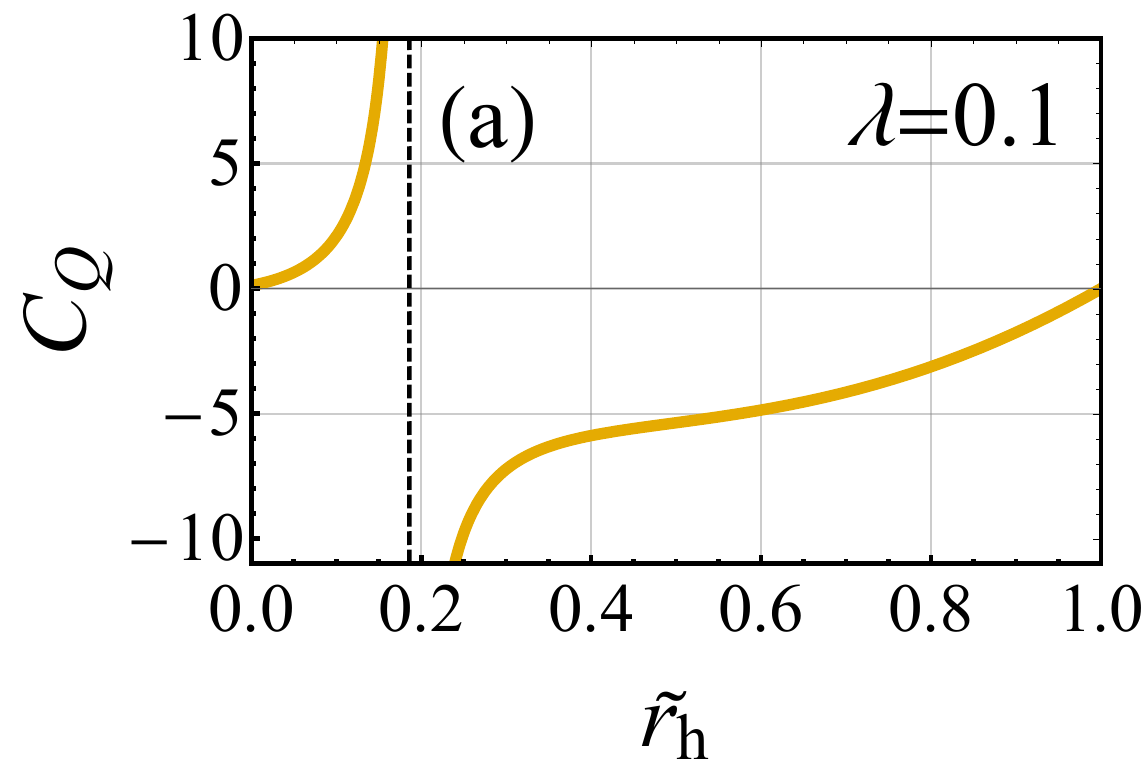}
\end{minipage}
}
\subfigure{\label{CQ2}
\begin{minipage}[t]{0.3\linewidth}
\centering
\includegraphics[width=1\textwidth]{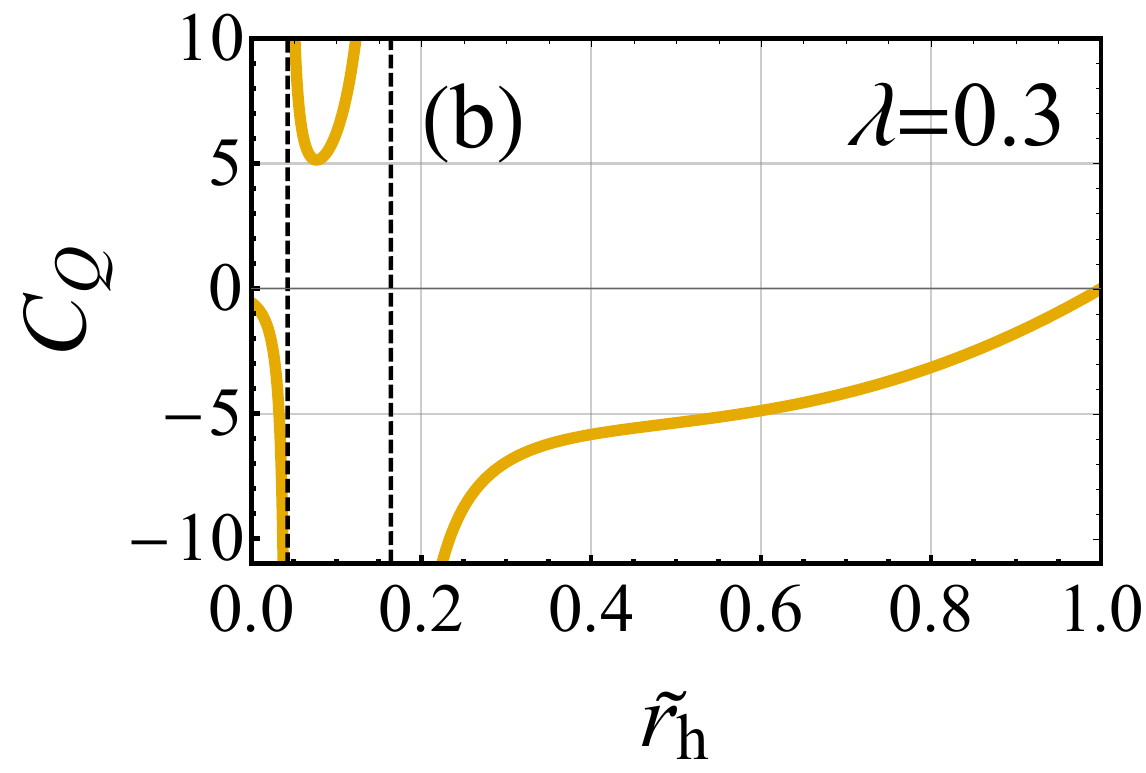}
\end{minipage}
}
\subfigure{\label{CQ3}
\begin{minipage}[t]{0.3\linewidth}
\centering
\includegraphics[width=1\textwidth]{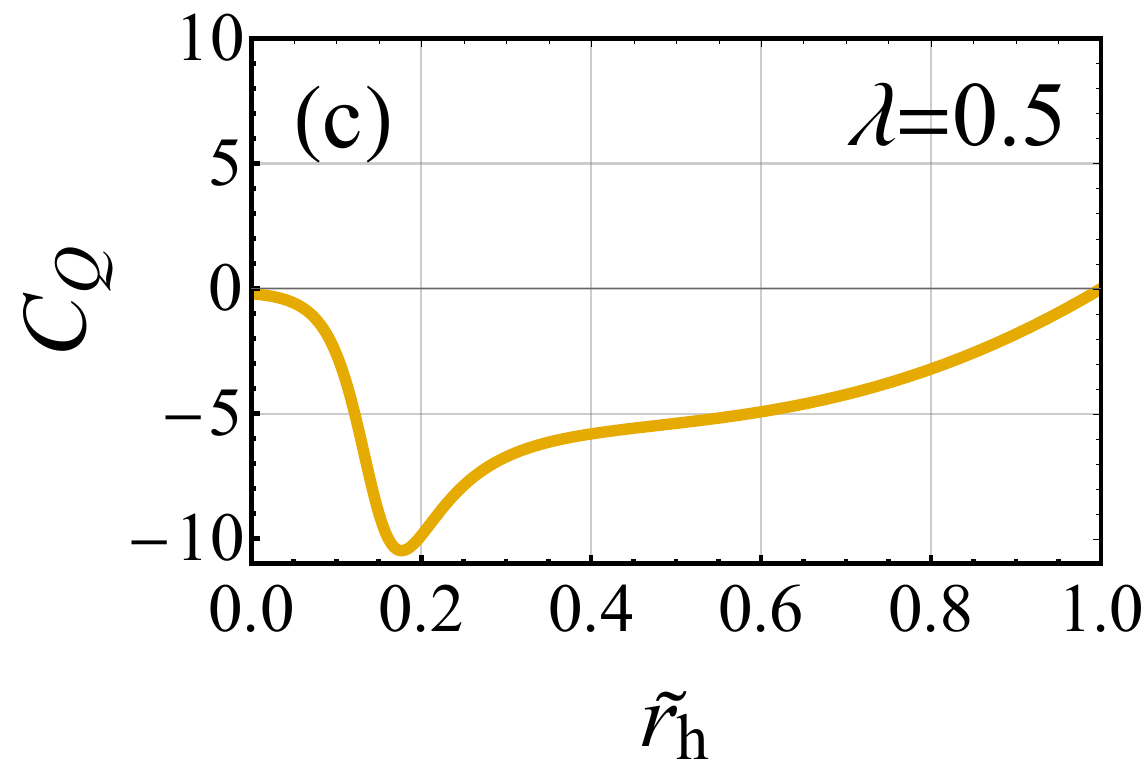}
\end{minipage}
}

\caption{Three distinct profiles of the heat capacity $C_Q$ as a function of the rescaled horizon radius $\tilde{r}_{\text{h}}$ under variation of the Euler-Heisenberg parameter $\lambda$, in which (a) $\lambda=0.1$, (b) $\lambda=0.3$ and (c) $\lambda=0.5$, while the other parameters are fixed as $R_0=1$, $Q=0.5$, $f_{R_0}=0.1$.}
\label{fig:CQlines}
\end{figure}

Fig. \ref{fig:CQlines} illustrates that the thermodynamic phase structures described by heat capacity $C_Q$ perform different behaviors under different black hole parameters. When the value of $\lambda$ is small, as shown in Fig. \ref{CQ1}, the nonlinear part of the electromagnetic field has little effect, and the heat capacity is primarily determined by the linear contribution, showing a structure similar to that of those simply charged black holes. As $\lambda$ increases, the influence of the nonlinear coupling terms becomes increasingly significant. A heat capacity structure with two divergence points emerges, as shown in Fig. \ref{CQ2}, and eventually, when the system is fully dominated by the nonlinear terms, a continuous and always negative heat capacity behavior appears, as shown in Fig. \ref{CQ3}. In other words, with the inclusion of the nonlinear coupling terms in Euler-Heisenberg theory and their increasing impact on the system, the black hole heat capacity progressively reveals two qualitatively distinct structural behaviors. Hence, we focus our study on the emergence of these new structures. Specifically, we compute the critical value of $\lambda$ at which the new structure appears between Fig. \ref{CQ1} and \ref{CQ2}, and take it as a reference point for the subsequent analysis. 

\section{QNMs of the $F(R)$-Euler-Heisenberg black holes}\label{sec3}

In this section, we adopt a different perspective and investigate the dynamical properties of the $F(R)$-Euler-Heisenberg black holes through the computation of QNMs. More precisely, we will focus on massless scalar field perturbations, with the numerical computation carried out through the pseudo-spectral method. 

From the general static and spherically symmetric metric Eq.(\ref{eq:met}), the Klein-Gordon equation can be derived as 
\begin{equation}
    \square \Psi = \frac{1}{\sqrt{-g}} \partial_\mu \left( \sqrt{-g} \, g^{\mu\nu} \, \partial_\nu \Psi \right) = 0.
\end{equation}
Performing the separation of variables $\Psi(t, r, \theta, \phi) = e^{-{\text{i}} \omega t} \, Y_{\ell m}(\theta, \phi) \, \frac{\psi(r)}{r}
$, we obtain
\begin{equation}
    \frac{\mathrm{d}^2 \psi}{\mathrm{d}r_*^2} + \left( \omega^2 - V_{\text{eff}}(r) \right) \psi = 0,\ \ V_{\text{eff}}(r) = h(r) \left( \frac{\ell(\ell+1)}{r^2} + \frac{h^{\prime}(r)}{r} \right),
    \label{eq: QNMs like-Sch}
\end{equation}
where $r_*$ is the tortoise coordinate  defined as $\text{d}r_*=\frac{\text{d}r}{h(r)}$, and $V_{\text{eff}}(r)$ is the effective potential.

By transforming into the ingoing Eddington-Finkelstein coordinate $\tilde{v}=t+r_*$ and applying a coordinate transformation $u=1/r$, the metric becomes 
\begin{equation}
    \mathrm{d}s^2=h(u)\mathrm{d}\tilde{v}^2-\frac{1}{u^2}\mathrm{d}\tilde{v}\mathrm{d}u+\frac{1}{u^2}(\mathrm{d}\theta^2+\mathrm{sin}^2\theta \mathrm{d}\phi^2),
\end{equation}
and the equation becomes
\begin{equation}
\left( \ell (1 + \ell) u + 2 {\text{i}} \omega \right)\, \psi+ \left( -2 {\text{i}} \omega u  - u^3  h^{\prime} \right)\, \psi^{\prime} - u^3 h\,   \psi^{\prime\prime}=0, 
\label{eq: QNMs eq}
\end{equation}
in which the prime $^{\prime}$ stands for $\frac{\mathrm{d}}{\mathrm{d}u}$, and the time-dependent coordinates are eliminated. 

Setting the black hole horizon and the cosmological horizon to be $u=u_{\text{b}}$ and $u=u_{\text{c}}$, the boundary conditions can be imposed in the pseudo-spectral method. Specifically, at the black hole horizon $u=u_{\text{b}}$, the ingoing boundary condition requires transforming the wave-like function $\psi$ into a special form such that the outgoing mode becomes divergent or highly oscillatory, making it unsuitable for pseudo-spectral representation, while the ingoing mode remains regular and can be accurately captured. A similar treatment applies at the cosmological horizon $u = u_{\text{c}}$ for enforcing the purely outgoing boundary condition. Under these conditions, the wave-like function $\psi$ must take the following form, 
\begin{equation}
   \psi= (u - u_{\text{c}})^{\left(-\frac{2 \mathrm{i} \omega}{u_{\text{c}}^2 h^{\prime}(u_{\text{c}})} - 
   1\right)}  \varphi,
\end{equation}
where $\varphi$ is a newly introduced undetermined function. Performing another coordinate transformation, $v=(u - u_{\text{c}})/(u_{\text{b}} - u_{\text{c}})$, by which the independent variable would be restricted to $v\in(0,1)$, we can finally obtain the final form of the equation, 
\begin{equation}
    A_0(v)\frac{\mathrm{d}^2\varphi}{\mathrm{d}v^2}+\Big(\mathrm{i}\omega B_1(v)+B_0(v)\Big)\frac{\mathrm{d}\varphi}{\mathrm{d}v}+\Big(\omega^2C_2(v)+\mathrm{i}\omega C_1(v)+C_0(v)\Big)\varphi=0,
    \label{eq:final eq}
\end{equation}
which is suitable for solving via the pseudo-spectral method, with the explicit forms of $A_0(v)$, $B_1(v)$, $B_0(v)$, $C_2(v)$, $C_1(v)$ and $C_0(v)$ provided in Appendix \ref{appendix2}. The corresponding QNMs can now be obtained by specifying the black hole parameters $R_0$, $Q$, $f_{R_0}$, $\lambda$, and $\tilde{r}_{\text{h}}$ and solving the equations accordingly. Specifically, we adopt the same prescription of rescaled black hole horizon radius in the analysis of QNMs as mentioned in the preceding discussion of thermodynamics, restricting the computation to those black holes that are thermodynamically well-defined and physically meaningful. {The numerical calculations follow the discussion in \cite{Jansen:2017oag}.}

\begin{figure}[H]
\centering
\subfigure{\label{o1}
\begin{minipage}[t]{0.4\linewidth}
\centering
\includegraphics[width=1\textwidth]{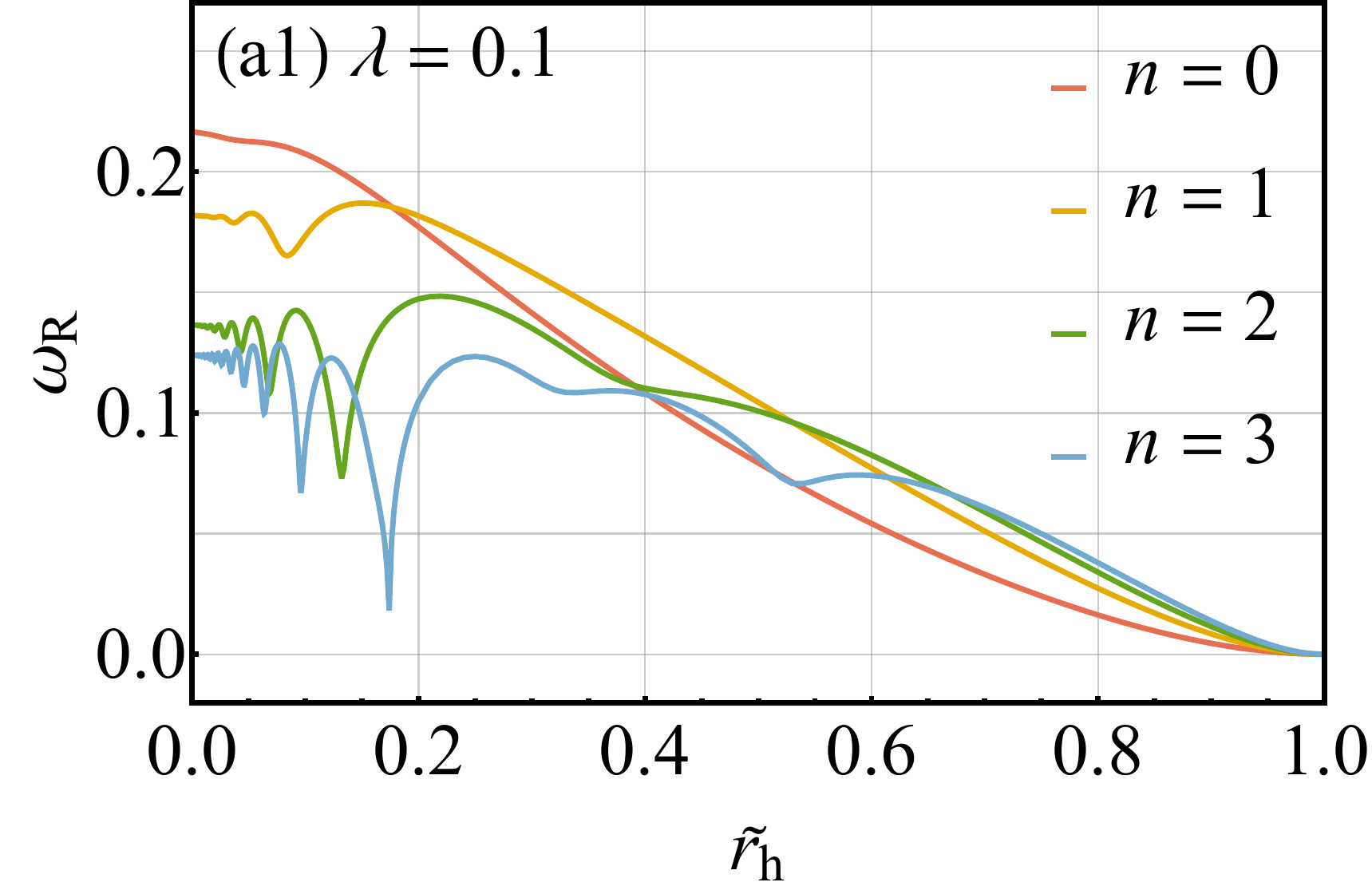}
\end{minipage}
\begin{minipage}[t]{0.4\linewidth}
\centering
\includegraphics[width=1\textwidth]{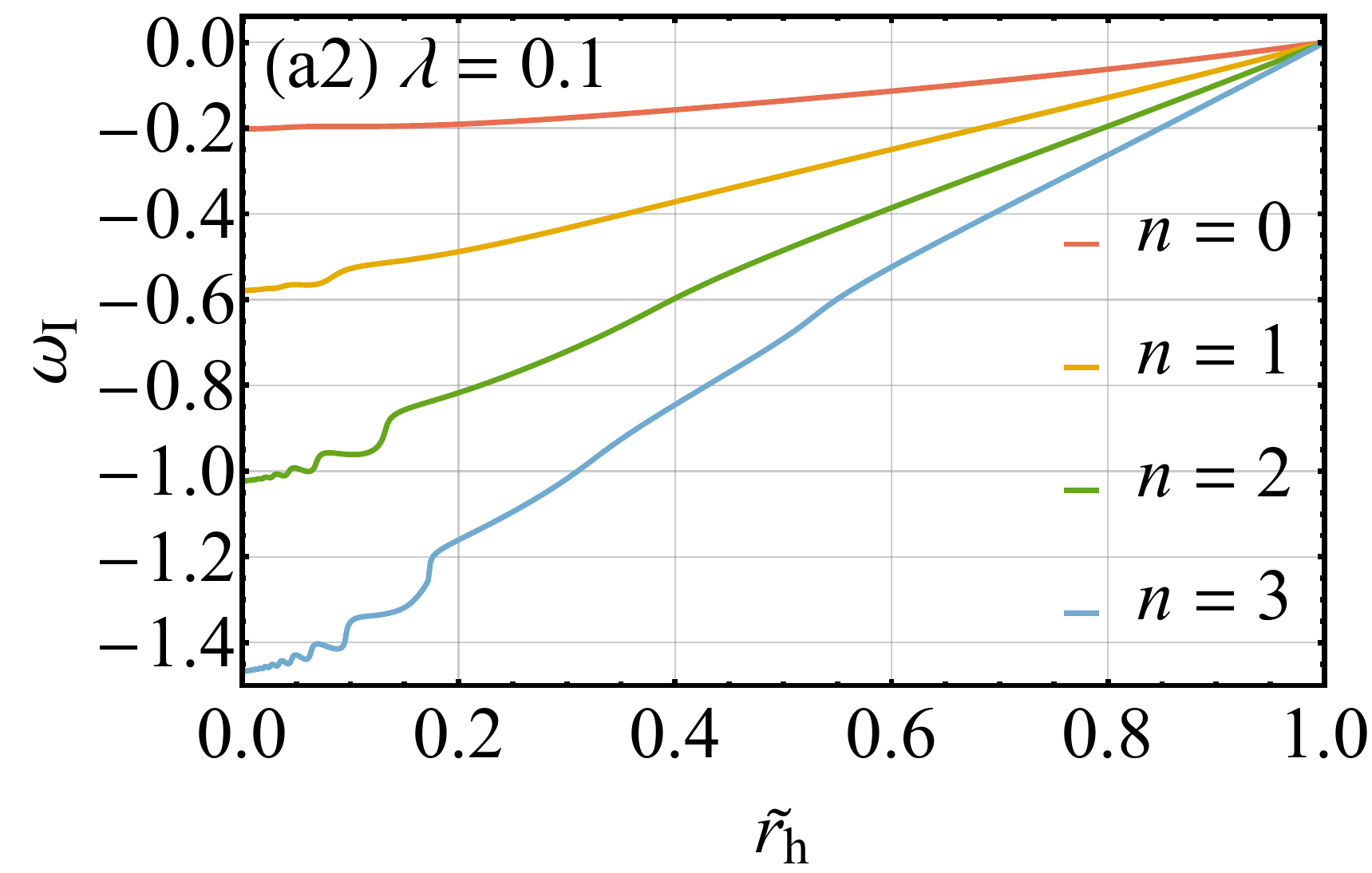}
\end{minipage}
}
\subfigure{\label{o2}
\begin{minipage}[t]{0.4\linewidth}
\centering
\includegraphics[width=1\textwidth]{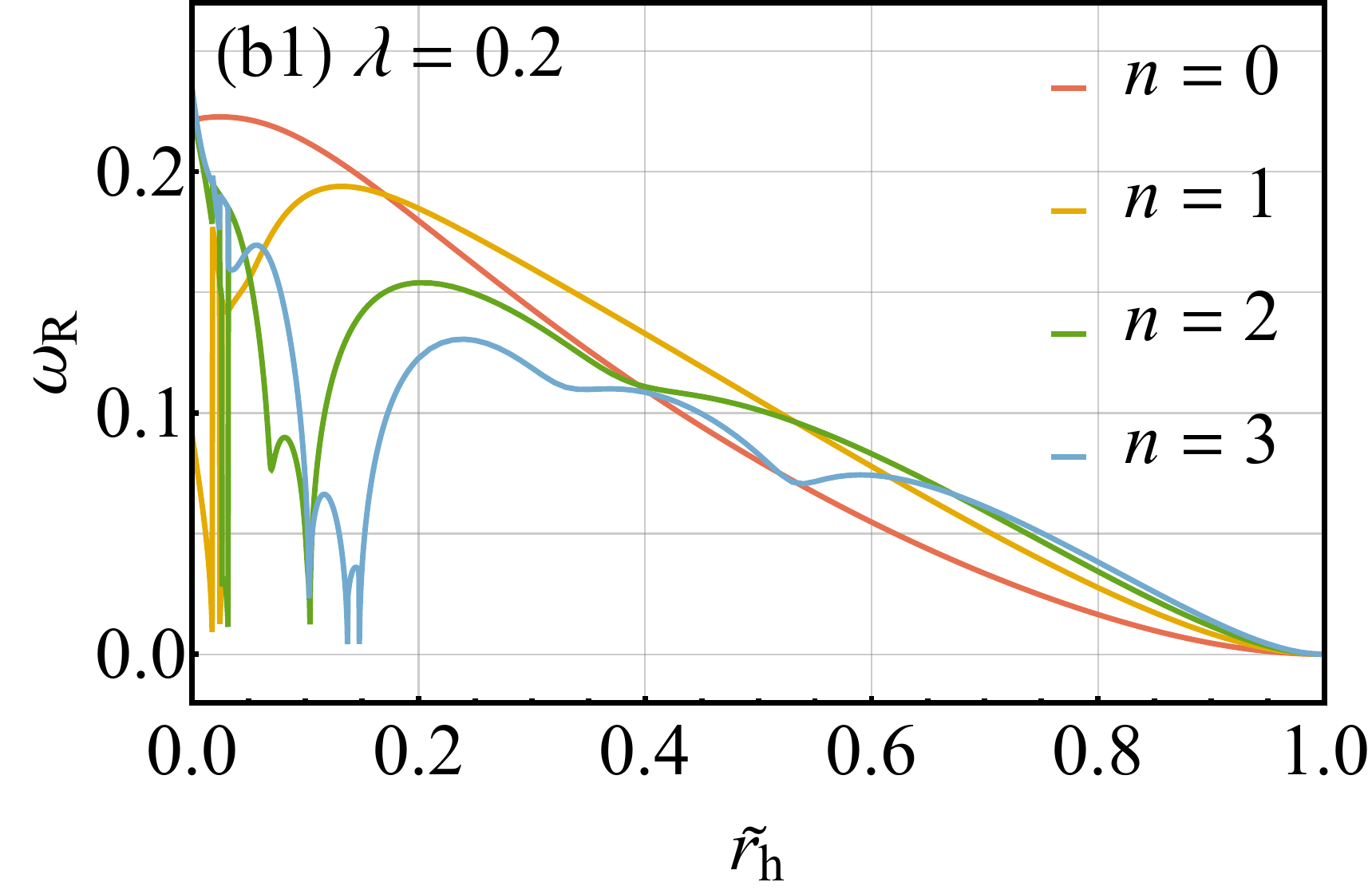}
\end{minipage}
\begin{minipage}[t]{0.4\linewidth}
\centering
\includegraphics[width=1\textwidth]{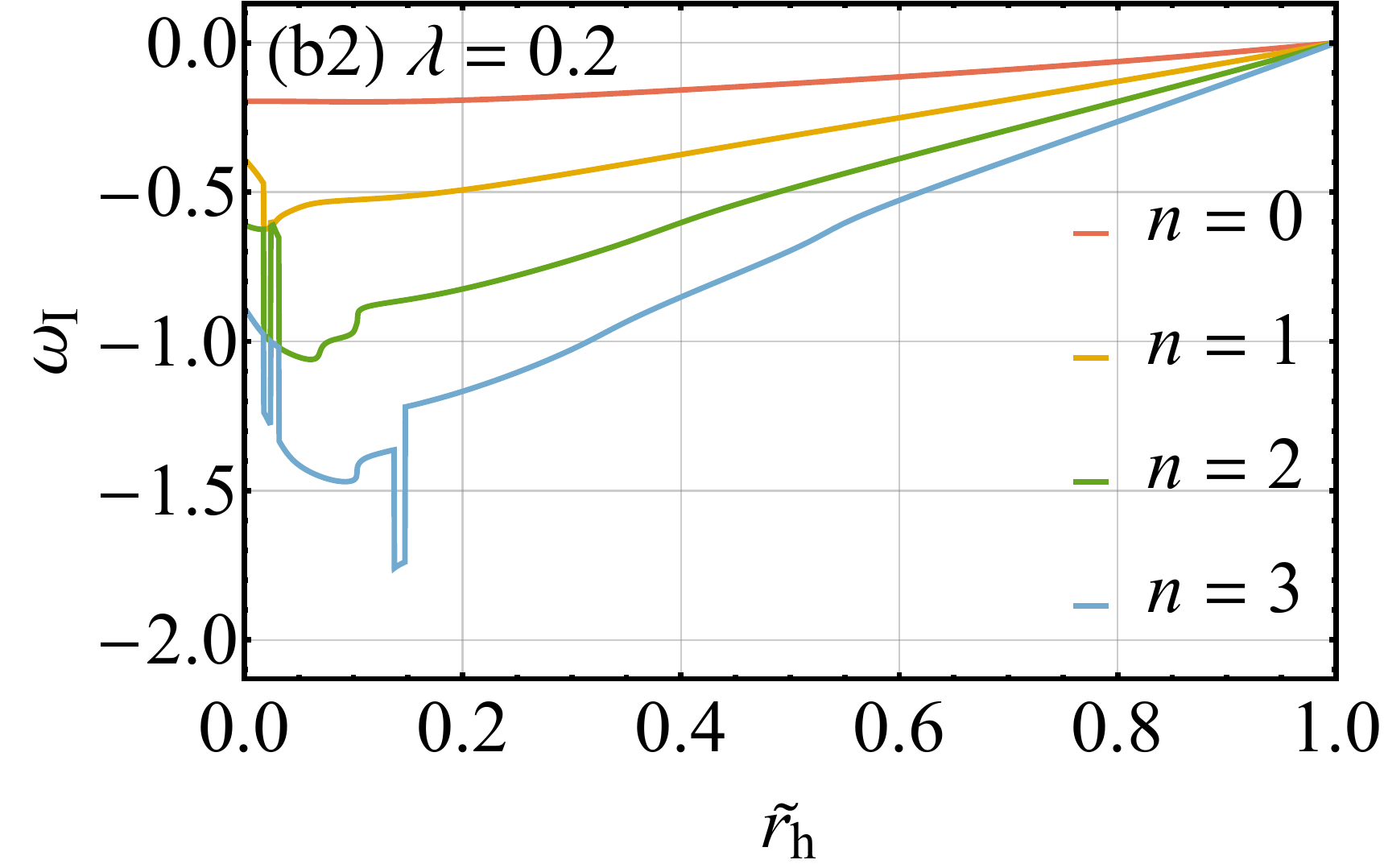}
\end{minipage}
}
\subfigure{\label{o3}
\begin{minipage}[t]{0.4\linewidth}
\centering
\includegraphics[width=1\textwidth]{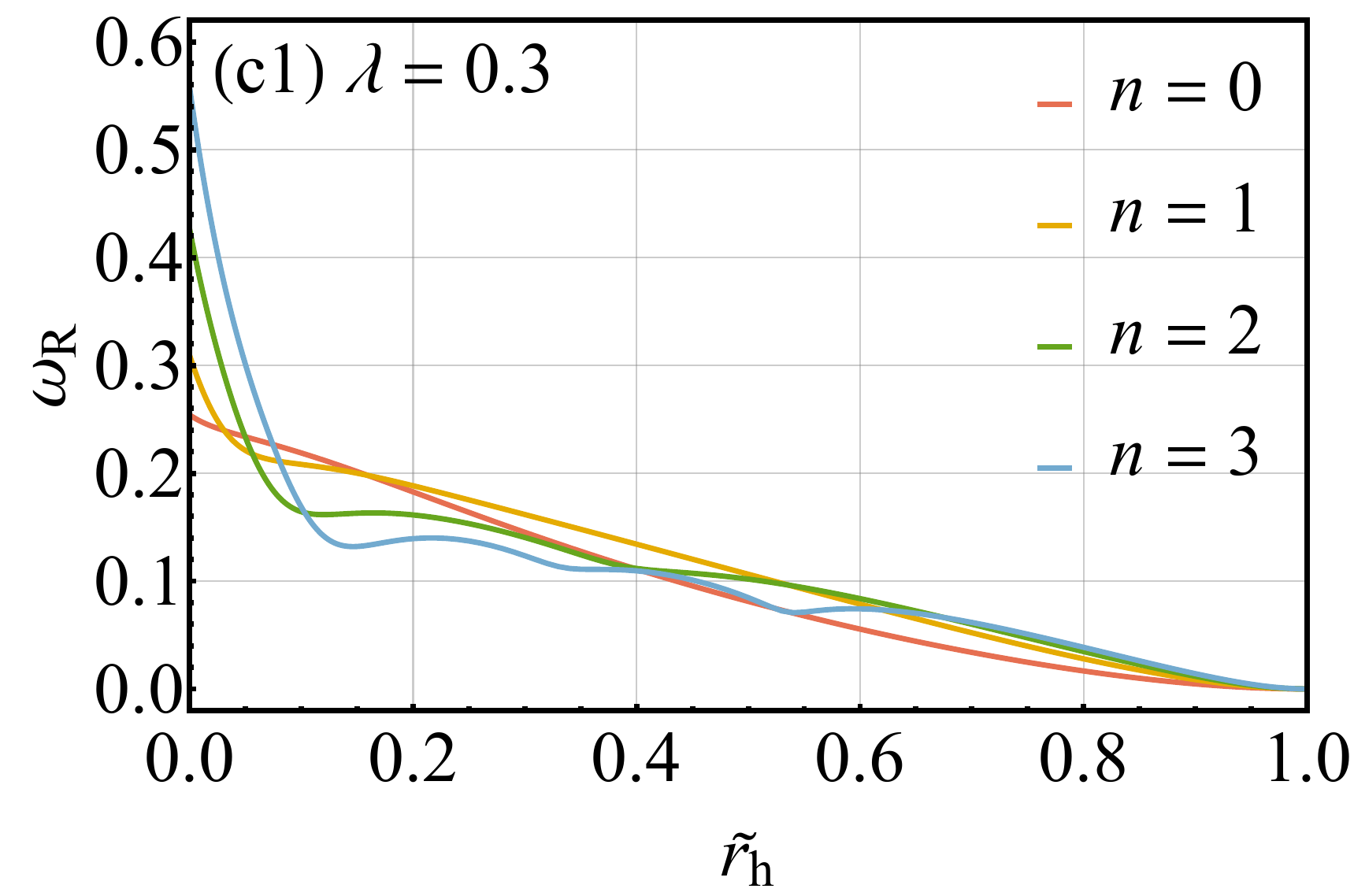}
\end{minipage}
\begin{minipage}[t]{0.4\linewidth}
\centering
\includegraphics[width=1\textwidth]{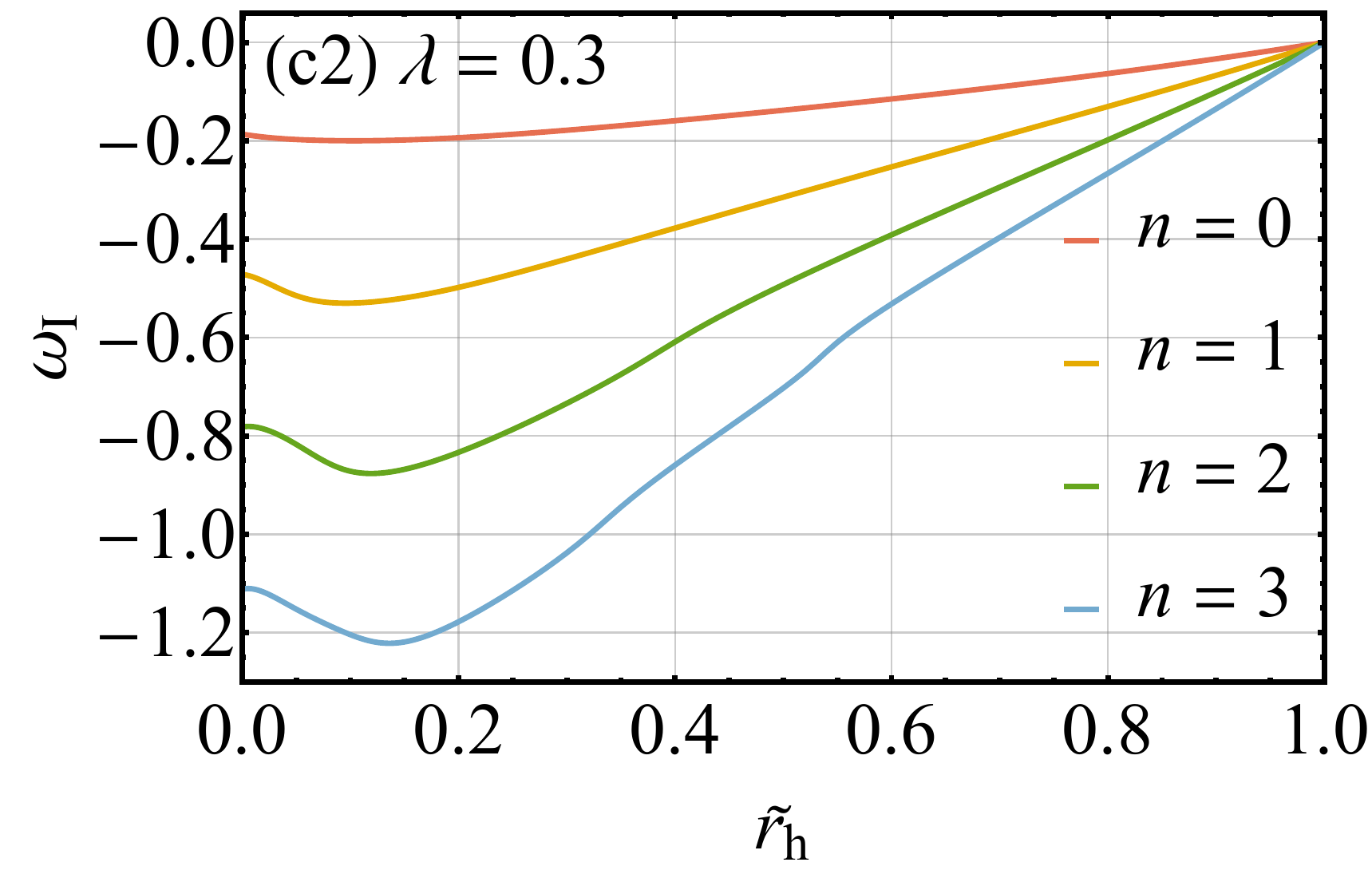}
\end{minipage}
}
    \caption{The real part $\omega_{\text{R}}$ and the imaginary part $\omega_{\text{I}}$ of the QNMs frequencies as a function of the rescaled horizon radius $\tilde{r}_{\text{h}}$ with $l=0$. The specific black hole parameter settings are (a) $\lambda=0.1$, (b) $\lambda=0.2$ and (c) $\lambda=0.3$, while the other parameters are fixed as $R_0=1,\ Q=0.5,\ f_{R_0}=0.1$. }
        \label{fig:omega to srh}
\end{figure}

Fig. \ref{fig:omega to srh} describes the QNMs for the case with zero angular quantum number ($l=0$) obtained by the pseudo-spectral method. The figures show the real part $\omega_{\text{R}}$ and the imaginary part $\omega_{\text{I}}$ of the QNMs frequencies versus the rescaled horizon radius $\tilde{r}_{\text{h}}$. The numerical results show that the imaginary parts of the QNMs frequencies are negative for the scalar field perturbation. That is to say, all these modes exhibit exponential decay over time, suggesting that $F(R)$-Euler-Heisenberg black hole remains stable under linear perturbations. Furthermore, it can also be observed that the absolute value of the imaginary part of the QNMs frequencies increases with the overtone number $n$. In other words, the higher the overtone, the faster the decay, whereas the fundamental mode ($n$ = 0) has the longest lifetime. These results all confirm the linear stability of the $F(R)$-Euler-Heisenberg black hole spacetime. 

We now analyze the effects induced by the variation of the Euler-Heisenberg parameter $\lambda$. For a small value of $\lambda$, the QNMs behave as shown in Fig. \ref{o1}. Specifically, when the overtone number equals or exceeds the critical value, namely $n \geq 1$, both the real and imaginary parts of the QNMs frequencies gradually become oscillatory functions as the black hole approaches the extremal limit $\tilde{r}_{\text{h}} = 0$. This result is fully consistent with the behavior of simply charged black holes (RN black holes \cite{PhysRevD.77.087501} and Kerr black holes \cite{Berti2003AsymptoticQM}). As $\lambda$ gradually increases, it can first be observed that, as shown in Fig. \ref{o2}, the oscillatory part develops a structure with dramatic changes,  which would not appear in the QNMs of other types of black holes. Such dramatic structural changes indicate that the QNMs spectrum undergoes a drastic global transformation at this stage. At the stage shown in Fig. \ref{o3}, the oscillatory behavior completely disappears. In other words, as the Euler-Heisenberg parameter $\lambda$ increases, the nonlinear electromagnetic effects begin to influence the near-extremal limit of the QNMs, leading to an overall change in the black hole dynamics as described by its QNMs. All of these phenomena depicted in the figures, structural transitions and newly emerging patterns, closely resemble those observed in the study of heat capacity.

In summary, for $F(R)$-Euler-Heisenberg black holes, there exist distinctive properties not only in the thermodynamics behavior of its heat capacity but also in the dynamical behavior of its QNMs.

\section{Analysis of the Consistency between QNMs and Heat Capacity}\label{sec4}

In the previous two sections, we have found that the $F(R)$-Euler-Heisenberg black holes exhibit novel structures in both their QNMs behavior and heat capacity structures. It is therefore worthwhile to investigate the possible correlation between the two aspects. 
{Introducing the slope parameter $K$ to characterize the variation of QNMs and taking the Euler-Heisenberg parameter $\lambda$ as a reference to compare the QNMs results with the corresponding thermodynamic behavior, we examine how changes in other black hole parameters affect this correspondence, and analyze the impact of varying the angular quantum number $l$ and the overtone number $n$ in the QNMs. }

We first focus on the zero mode ($n = 0$) with the angular quantum number taking $l=0$ and set a \textbf{slope parameter} $K $ \cite{Jing:2008an}, taking $\tilde{r}_{\text{h}}$ as the implicit variable, 
\begin{equation}
    K = \frac{\mathrm{d}\omega_{\text{I}}/\mathrm{d}\tilde{r}_{\text{h}}}{\mathrm{d}\omega_{\text{R}}/\mathrm{d}\tilde{r}_{\text{h}}}.
\end{equation}

\begin{figure}[H]
\ 

  \centering
\subfigure{\label{kr1} \begin{minipage}[t]{0.45\textwidth}
    \centering
    \includegraphics[width=\textwidth]{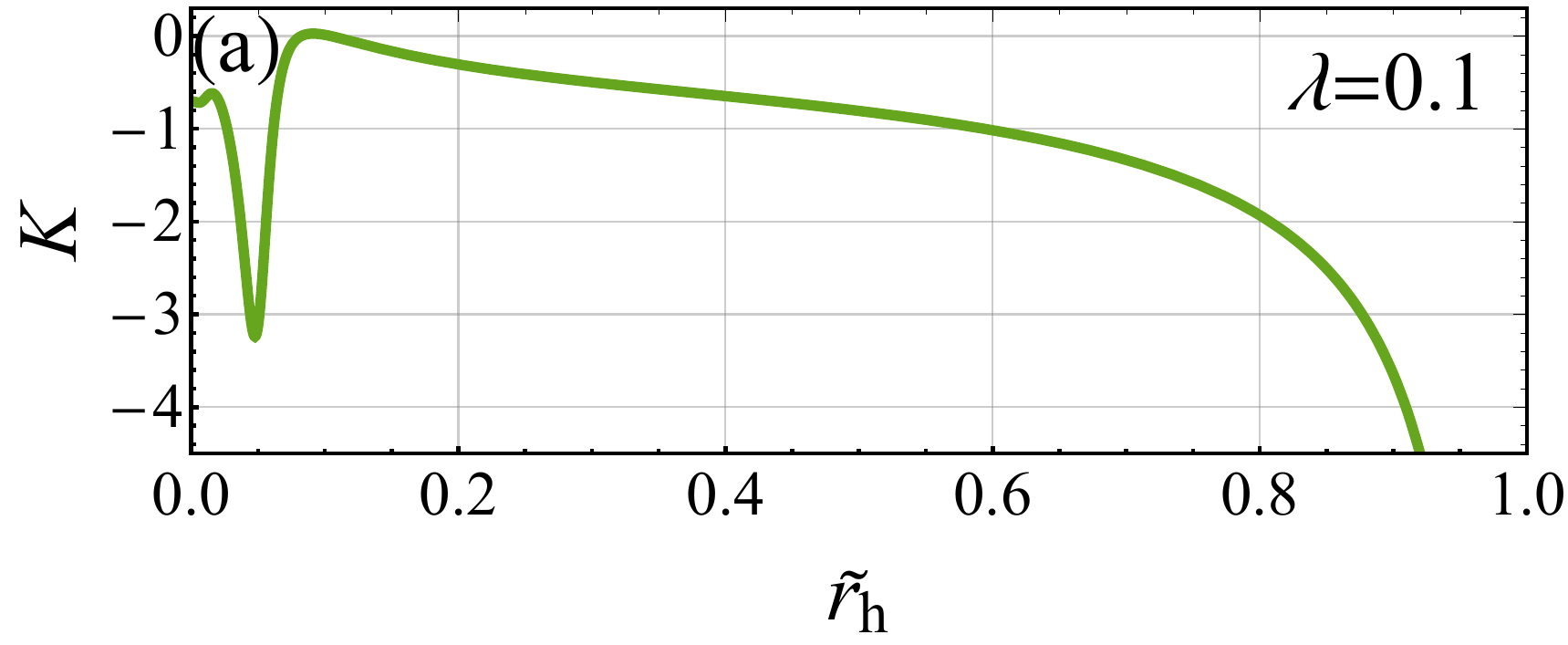}
  \end{minipage} }

\subfigure{\label{kr2} \begin{minipage}[t]{0.45\textwidth}
    \centering
    \includegraphics[width=\textwidth]{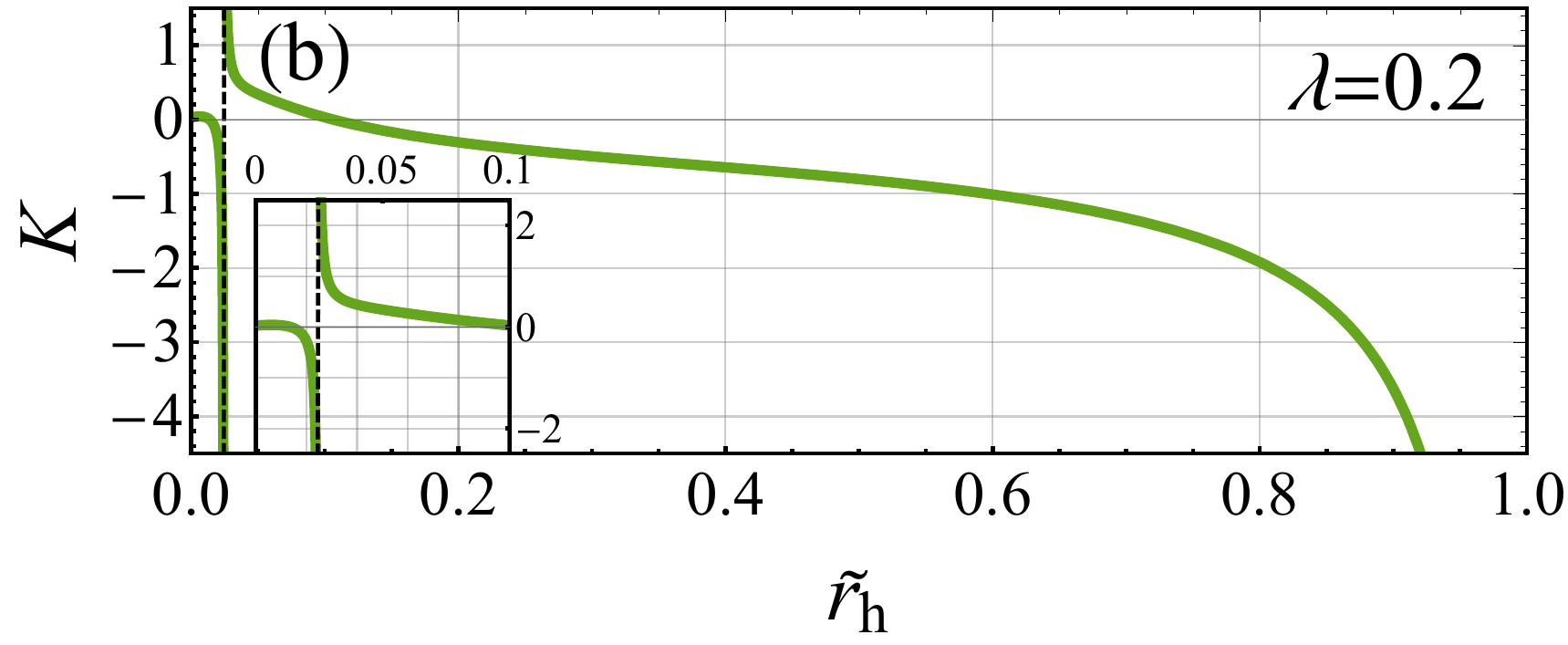}
  \end{minipage} }

  \centering
\subfigure{\label{kr3} \begin{minipage}[t]{0.45\textwidth}
    \centering
    \includegraphics[width=\textwidth]{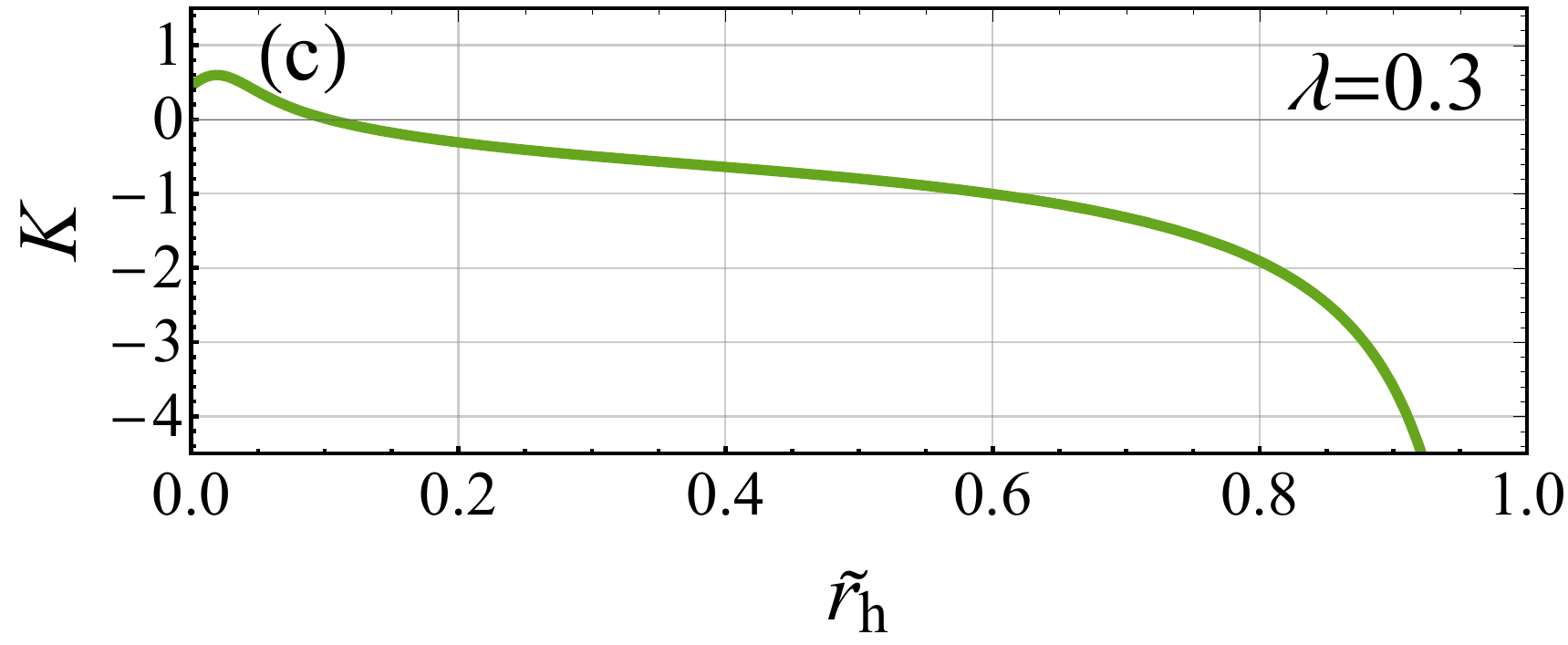}
  \end{minipage} }

 \caption{Three distinct profiles of the QNMs frequencies slope $K = \frac{\mathrm{d}\omega_{\text{I}}/\mathrm{d}\tilde{r}_{\text{h}}}{\mathrm{d}\omega_{\text{R}}/\mathrm{d}\tilde{r}_{\text{h}}}$ as a function of the rescaled horizon radius $\tilde{r_h}$ under variation of $\lambda$. Specifically, the panels show the full range $[0,1]$ of $\tilde{r}_{\text{h}}$ , while the small panel in (b) corresponds to the region $[0,0.1]$, where structural transitions occur. The detail parameter settings are $R_0=1,$ $ Q=0.5$, $f_{R_0}=0.1$, and (a) $\lambda=0.1$, (b) $\lambda=0.2$, (c) $\lambda=0.3$. }
  \label{fig:QNMslines}
\end{figure}

Similarly to the analysis of the heat capacity, one can draw the slope parameter $K$ versus the rescaled horizon radius $\tilde{r}_{\text{h}}$ and investigate the variation of its structure by increasing the Euler-Heisenberg parameter $\lambda$ in the same way. As shown in Fig. \ref{fig:QNMslines},  structural differences emerge near the limit $\tilde{r}_{\text{h}}=0$. Specifically, there are three distinct structures, one with a single divergence point with the other two types of structures on either side of it. 

Interestingly, if we calculate the critical value of the black hole parameters where the transitions between different structures occur, one can find a notable numerical match. To be more precise, the transition from the structures of heat capacity between Fig. \ref{CQ1} and \ref{CQ2} occurs at $\lambda=0.225$ with the other black hole parameters taking the values $R_0=1$, $Q=0.5$ and $f_{R_0}=0.1$; at the same time, the transition from the structures of QNMs between Fig. \ref{kr2} and \ref{kr3} takes place at $\lambda=0.232$ under the same black hole parameter settings. It follows that the critical value of $\lambda$ at which one of the phase structure transitions occurs is nearly equal to the critical value at which one of the structural transitions takes place in the QNMs, moreover, the discrepancy is only $3.4\%$ and could be regarded as computational error, assuming that the agreement is not simply due to a numerical coincidence. 

\subsection{Relative Error Parameter}

\begin{figure}[ht!]
  \centering
  \includegraphics[width=0.5\linewidth]{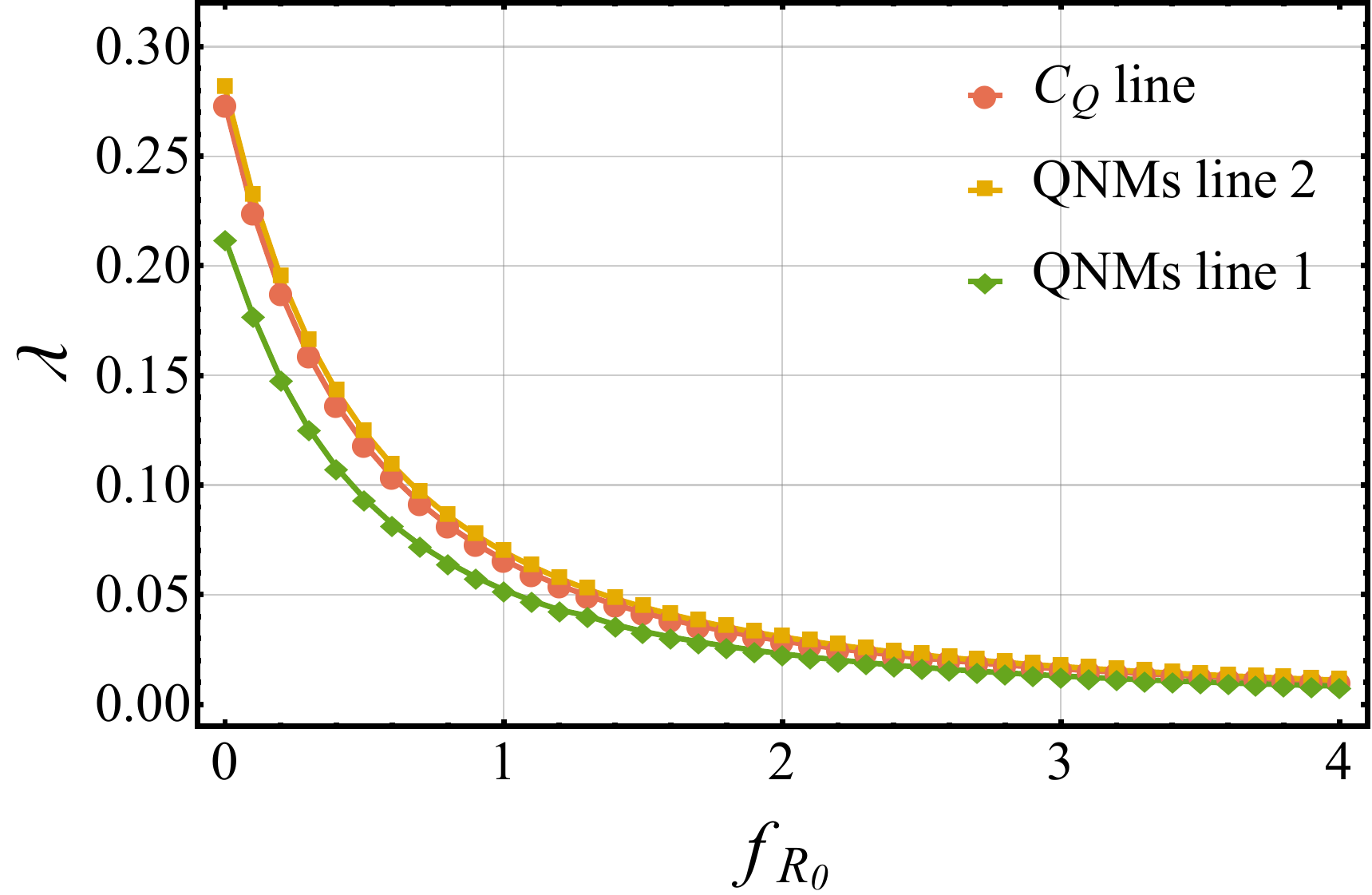}
  \hfill
 \caption{Three kinds of structure transitions under the parameters setting $R_0=1$ and $Q=1/2$, in which the parameter $f_{R_0}$ is varied from 0 to a value where the corresponding critical $\lambda$ becomes sufficiently close to zero and its variation becomes negligible, specifically, 41 data points within the range from 0 to 4 are selected as reference values. The different curves represent the critical values of $\lambda$ at which different structural transitions occur, for heat capacity and QNMs.}
  \label{fig:CQcomp}
\end{figure}

The computation can be extended to the entire $\lambda -f_{R_0}$ plane with fixed values of $Q$ and $R_0$. We still take the Euler-Heisenberg parameter $\lambda$ as the comparison quantity and compute the critical values of $\lambda$ corresponding to each transition by fixing different values of $f_{R_0}$. As shown in Fig. \ref{fig:CQcomp}, there are two critical curves derived independently from the QNMs and the heat capacity computations nearly coincide, indicating a concrete consistency between them. Specifically, ``QNMs line'' and ``$C_Q$ line'' respectively represent the structure transitions observed in the analyses of the QNMs and the black hole heat capacities, and the number ``1'' and ``2'' represent the first and second transitions as the value of Euler-Heisenberg parameter $\lambda$ increases, respectively.

To further investigate the consistency, we set a \textbf{relative error parameter} to be 
\begin{equation}
    \Delta_\lambda = \left(\lambda_{\text{C.QNMs}}-\lambda_{\text{C.}{C_Q}}\right)/\lambda_{\text{C.}{C_Q}}, 
\end{equation}
in which $\lambda_{\text{C.QNMs}}$ and $\lambda_{\text{C.}{C_Q}}$ represent the related critical values of $\lambda$ to the transitions in the QNMs and the heat capacities under fixed choices of other parameters, respectively. In this way, we are able to analyze the effects on this consistency of all other black hole parameters, including $R_0$, $Q$ and $f_{R_0}$.
\ 

\begin{figure}[!ht]
    \centering
  \subfigure{\label{changeR0} \begin{minipage}[t]{0.45\textwidth}
    \centering
    \includegraphics[width=\textwidth]{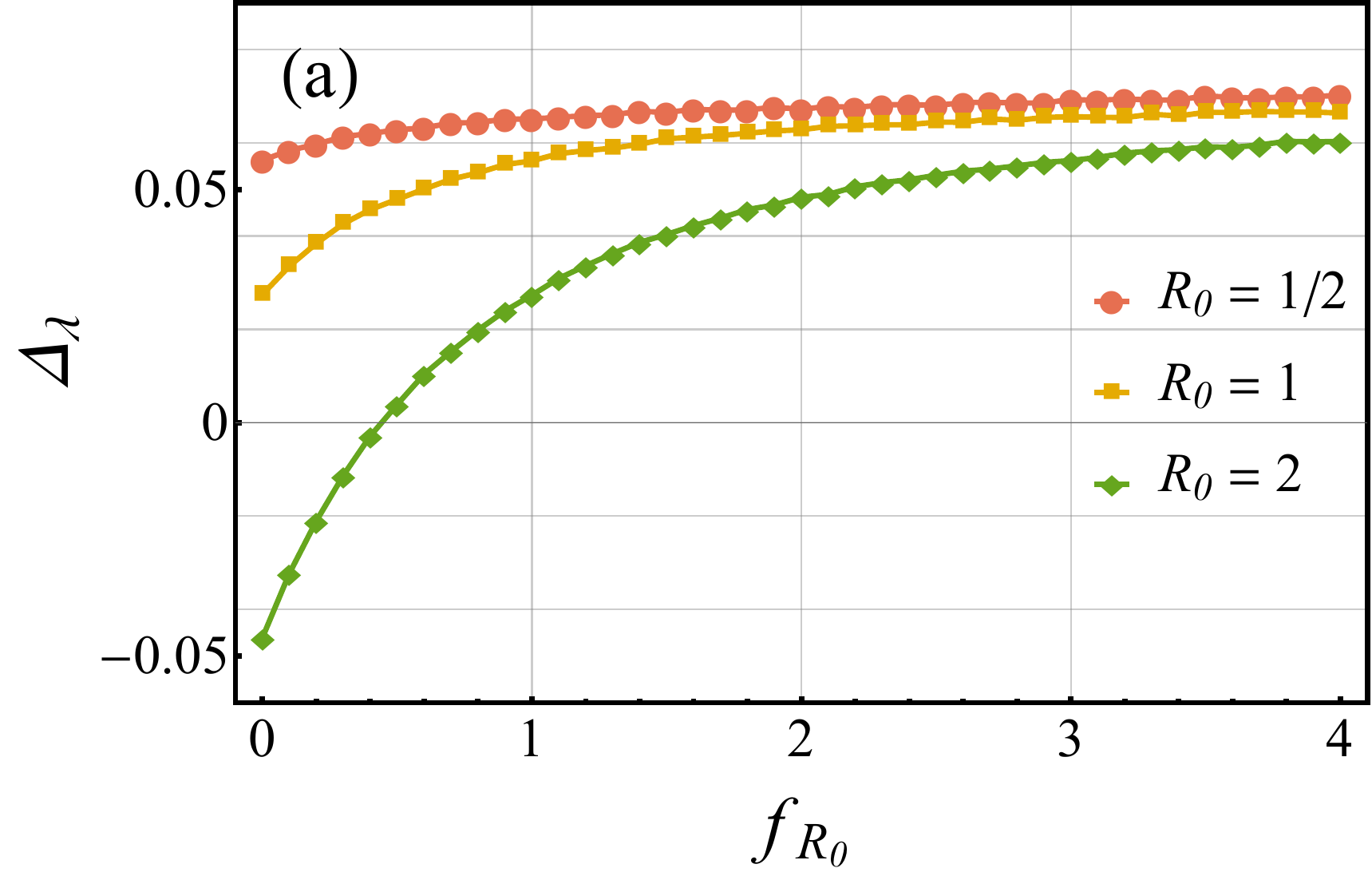}
  \end{minipage} }
  \hfill
\subfigure{\label{changeQ} \begin{minipage}[t]{0.45\textwidth}
    \centering
    \includegraphics[width=\textwidth]{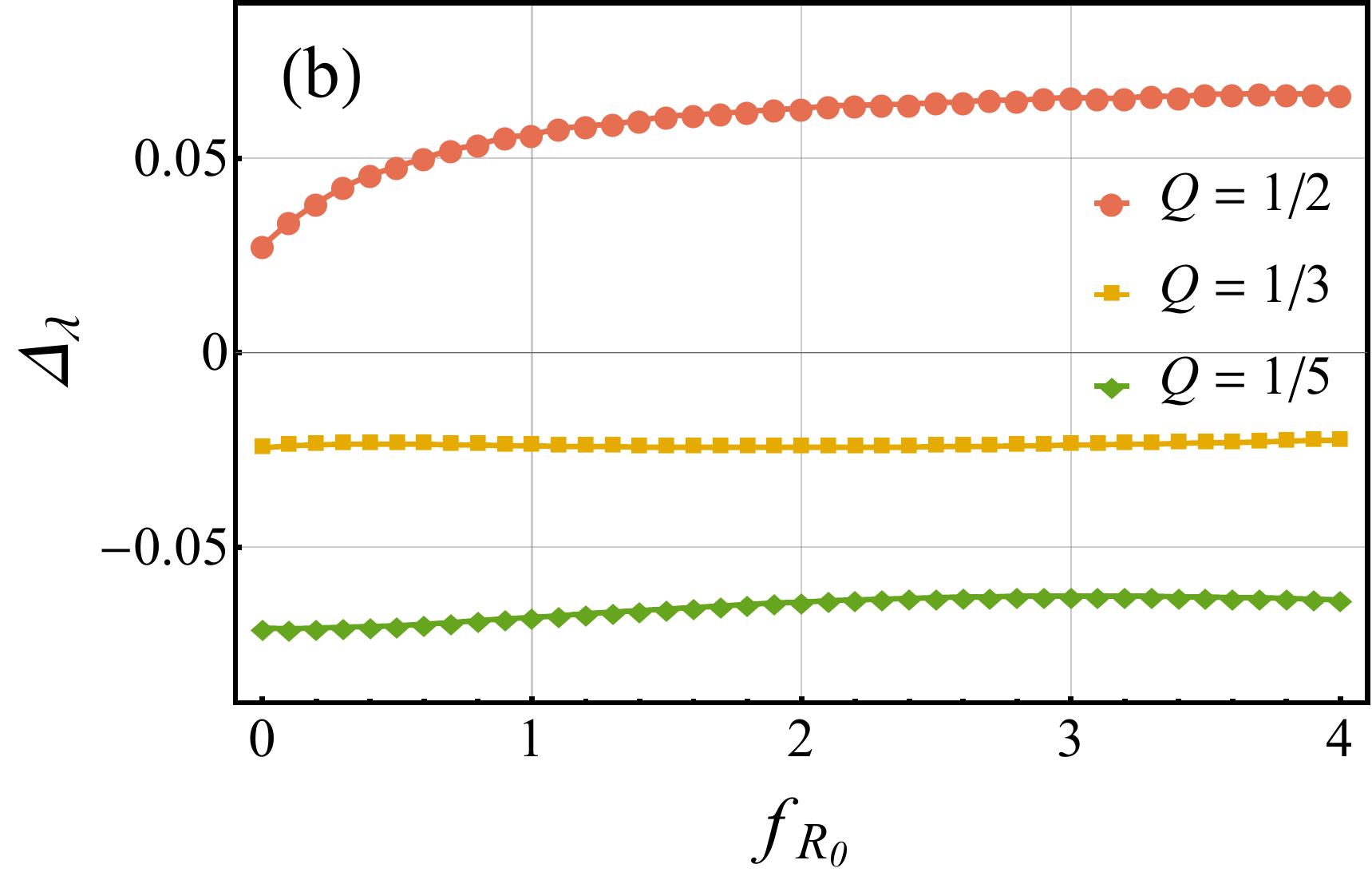}
  \end{minipage} }
  \hfill
    \caption{
The influences on the relative error $\Delta_\lambda$ as a function of $f_{R_0}$ of varying the parameter (a) $R_0$ with fixed $Q = \frac{1}{2}$ and (b) $Q$ with fixed $R_0 = 1$. The sign of the relative error $\Delta_\lambda$ reflects the relationship between the two critical values. Specifically, $\Delta_\lambda>0$ indicates that $\lambda_{\text{C.QNMs}}>\lambda_{\text{C.}C_Q}$, whereas $\Delta_\lambda<0$ implies $\lambda_{\text{C.QNMs}}<\lambda_{\text{C.}C_Q}$.}
    \label{fig:R0 and Q}
\end{figure}

{It is demonstrated in Fig. \ref{fig:R0 and Q} how the relative error $\Delta_\lambda$ responds to changes in the black hole parameters $f_{R_0}$, $R_0$ and $Q$, confirming that the universal correspondence remains robust across different parameter choices.} Each individual curve in the figure describes the effect on the relative error $\Delta_\lambda$ of the parameter $f_{R_0}$, while their combination futher reveals the effects of the parameters $R_0$ and $Q$, respectively. It can be seen that the value of $\Delta_\lambda$ may take positive or negative values in different cases, implying that there is no strict ordering between the two critical values of $\lambda$. Specifically, for varying $f_{R_0}$, the value of $\Delta_\lambda$ increases with increasing $f_{R_0}$ and gradually approaches a stable limit; for varying $R_0$, the value of $\Delta_\lambda$ increases with decreasing $R_0$; and for varying $Q$, the value of $\Delta_\lambda$ increases with increasing $Q$. Actually, based on the Eq.(\ref{eq:metric func}) and Eq.(\ref{eq:mass perfrom}), these variations, including an increase in $f_{R_0}$, a decrease in $R_0$ or an increase in $Q$, also result in another effect, a contraction of the parameter space of $\lambda$, or in other words, a decrease in all critical values of $\lambda$. Consequently, the relative error $\Delta_\lambda$ is essentially determined by the scaling of the parameter space of $\lambda$, or the magnitude of the critical value of $\lambda$, although it originates from variations in different black hole parameters.

\ 

\subsection{Angular and Overtone Number Effects}

In some recent studies, quantum effects in gravitational theories have been found to be obscured by increasing the angular quantum number of QNMs \cite{Song:2024kkx}. We now consider the influence of the angular quantum number $l$ and the overtone number $n$ on the QNMs, and begin with different overtone numbers $n$ with the angular quantum number fixed in $l=0$. 

\begin{figure}[H]

  \centering  
\subfigure{\label{n11} \begin{minipage}[t]{0.3\textwidth}
    \centering
    \includegraphics[width=\textwidth]{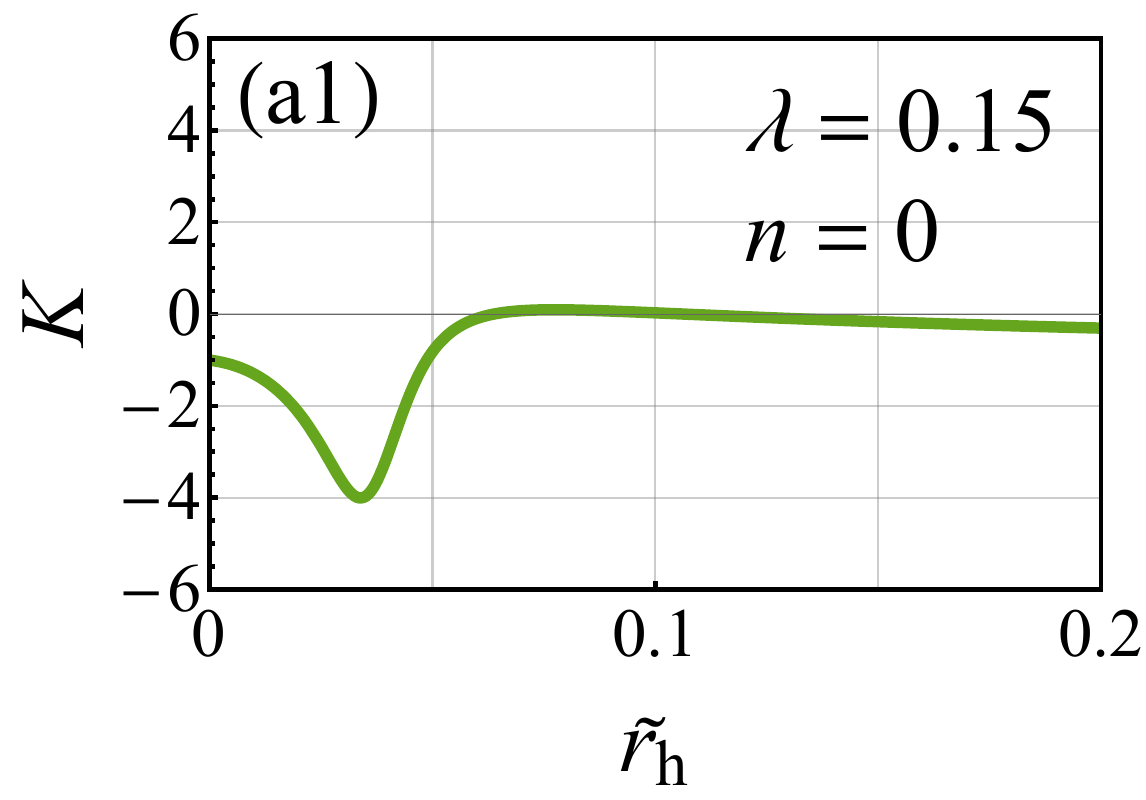}
  \end{minipage} }
  \hfill
\subfigure{\label{n12} \begin{minipage}[t]{0.3\textwidth}
    \centering
    \includegraphics[width=\textwidth]{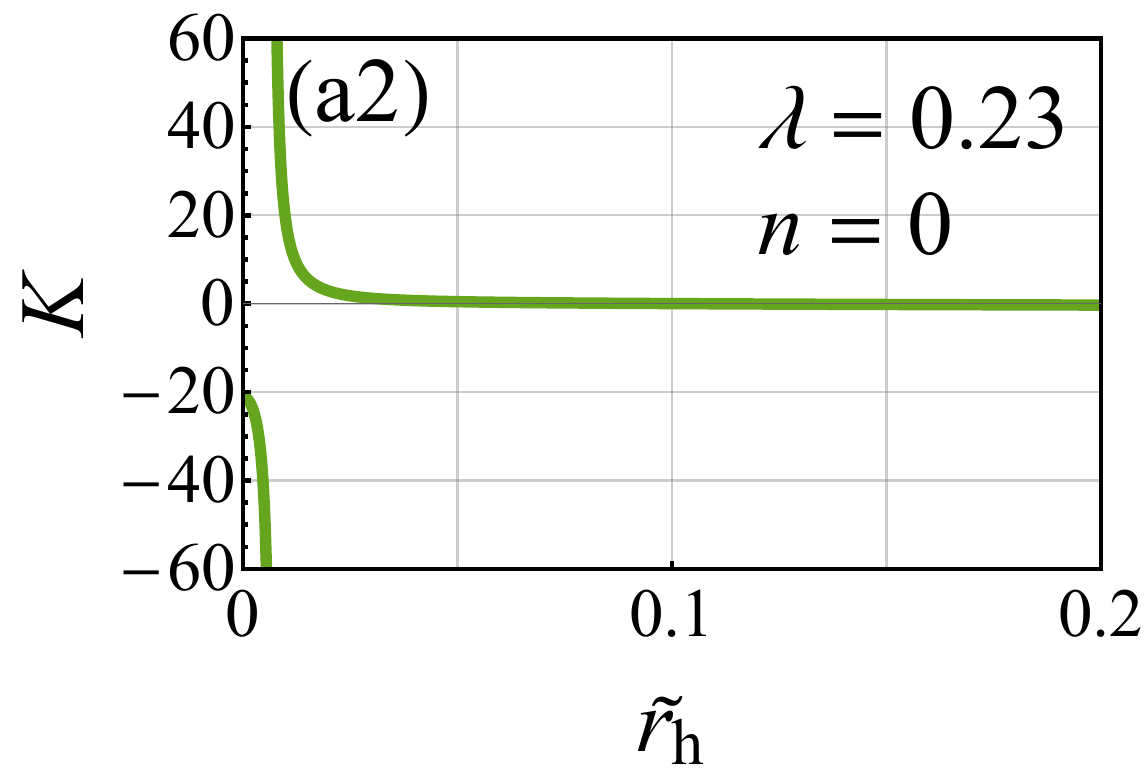}
  \end{minipage} }
  \hfill
\subfigure{\label{n13} \begin{minipage}[t]{0.3\textwidth}
    \centering
    \includegraphics[width=\textwidth]{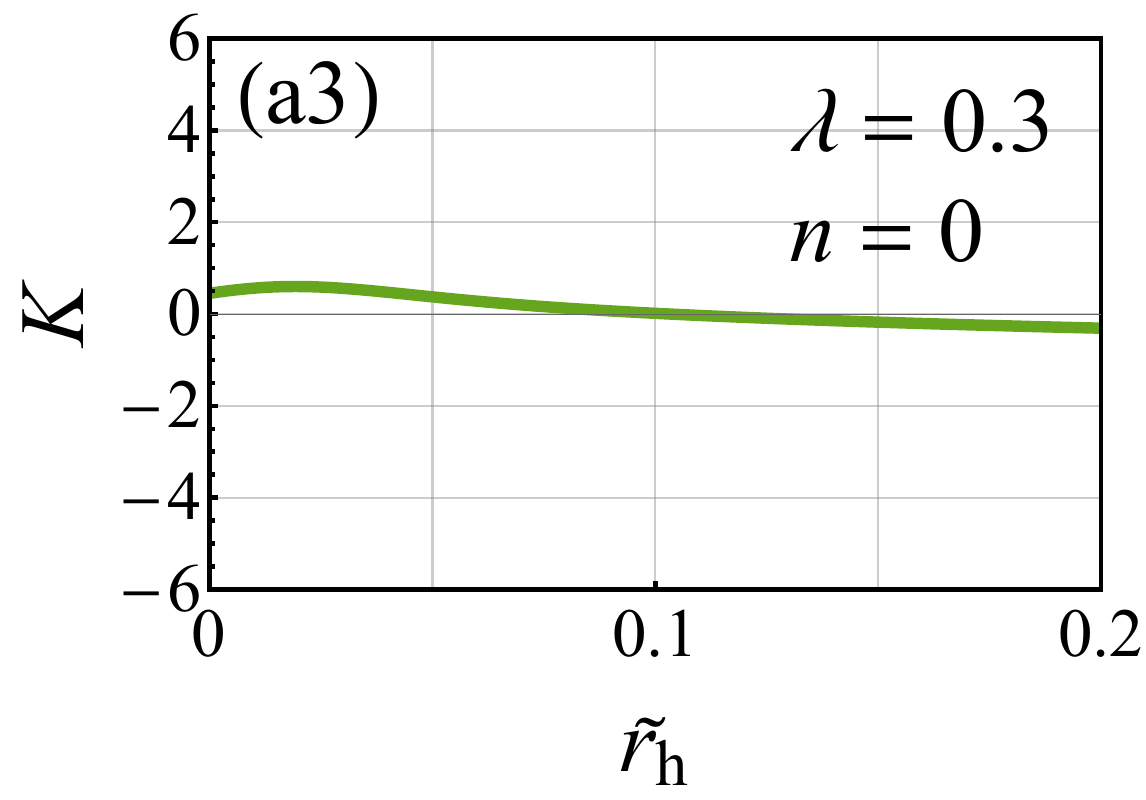}
  \end{minipage} }

  \centering  
\subfigure{\label{n21} \begin{minipage}[t]{0.3\textwidth}
    \centering
    \includegraphics[width=\textwidth]{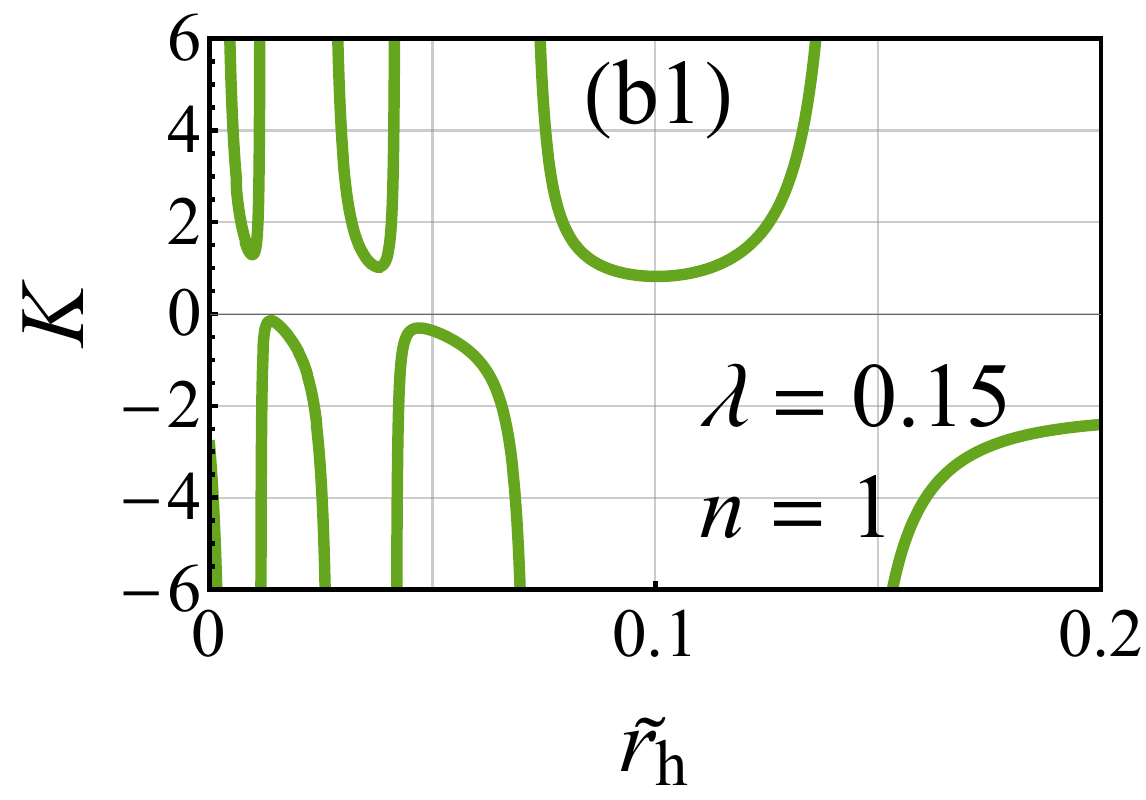}
  \end{minipage} }
  \hfill
\subfigure{\label{n22} \begin{minipage}[t]{0.3\textwidth}
    \centering
    \includegraphics[width=\textwidth]{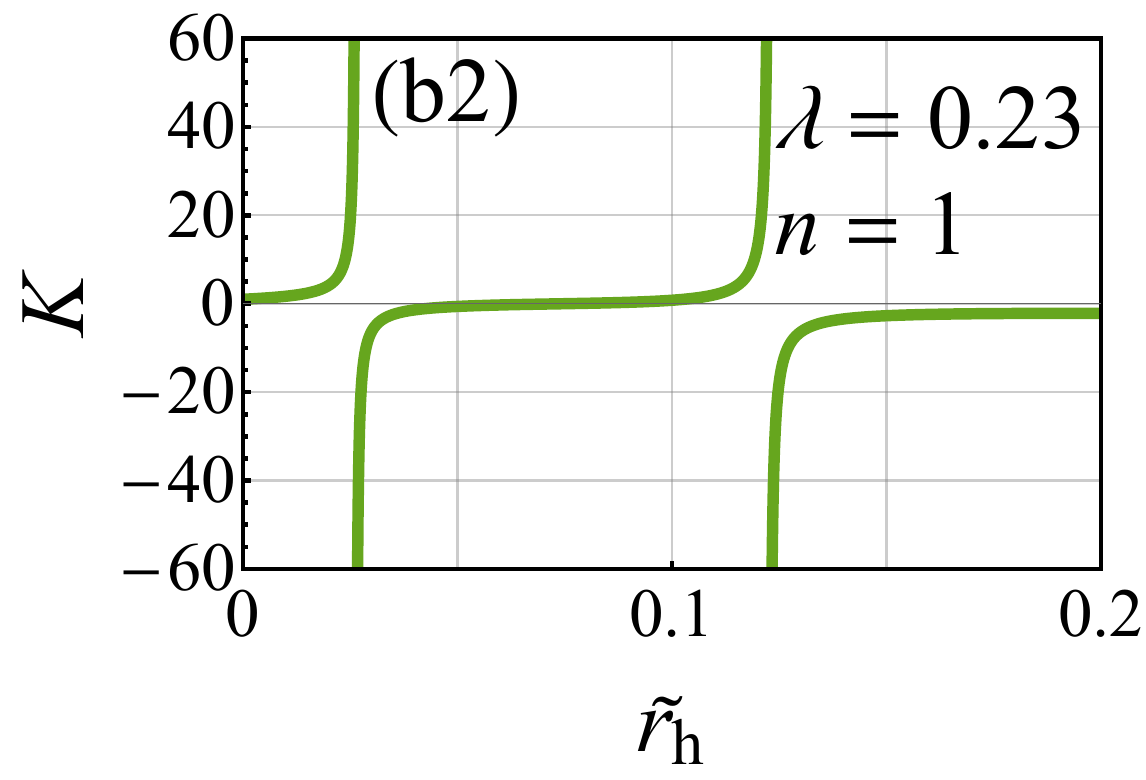}
  \end{minipage} }
  \hfill
\subfigure{\label{n23} \begin{minipage}[t]{0.3\textwidth}
    \centering
    \includegraphics[width=\textwidth]{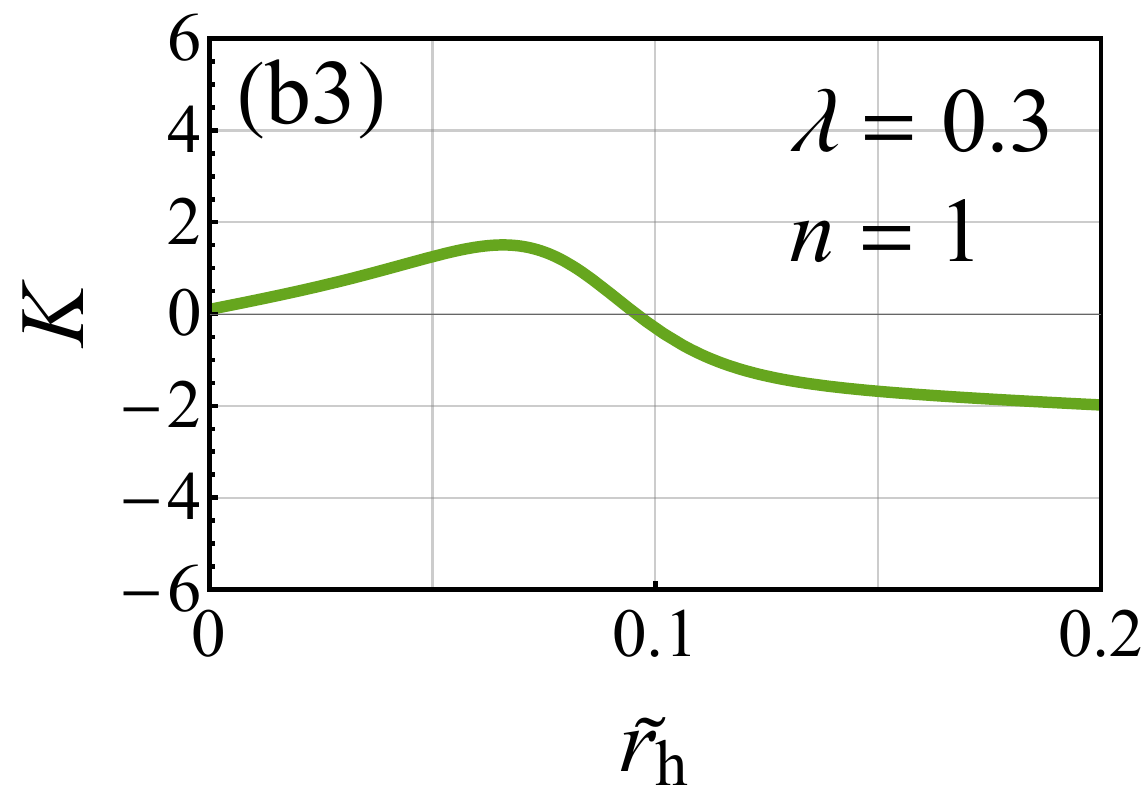}
  \end{minipage} }

  \centering  
\subfigure{\label{n31} \begin{minipage}[t]{0.3\textwidth}
    \centering
    \includegraphics[width=\textwidth]{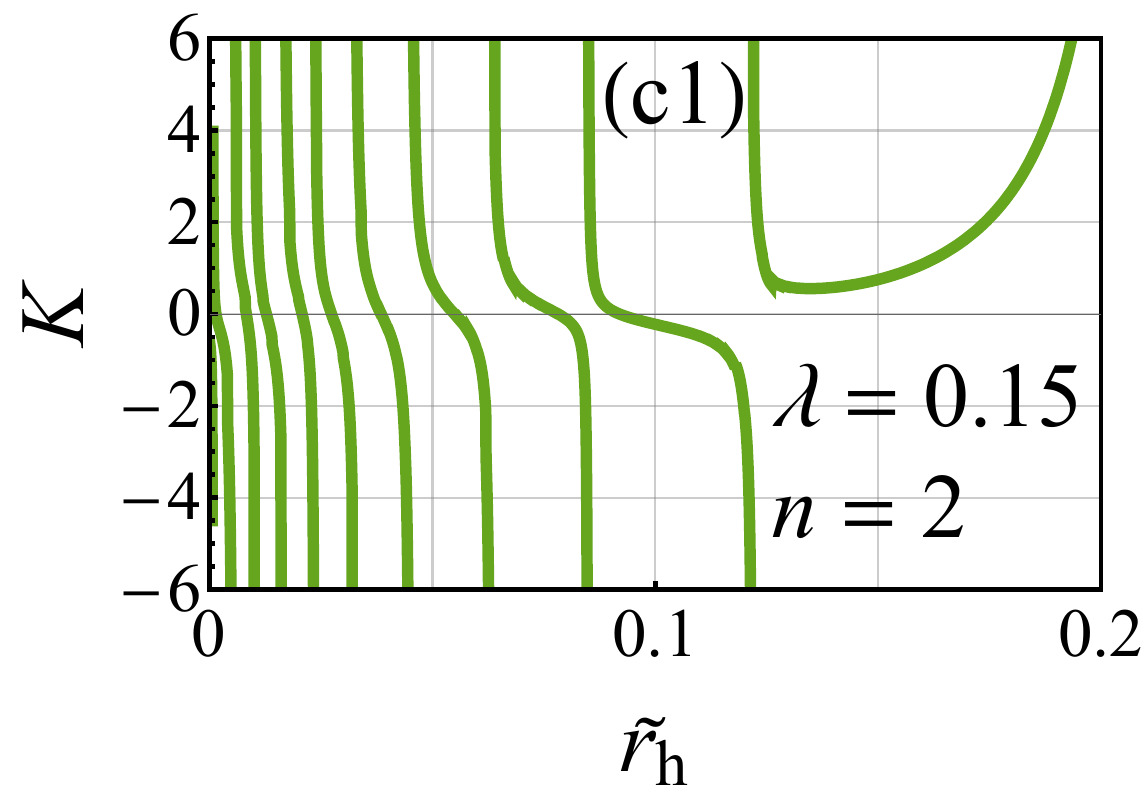}
  \end{minipage} }
  \hfill
\subfigure{\label{n32} \begin{minipage}[t]{0.3\textwidth}
    \centering
    \includegraphics[width=\textwidth]{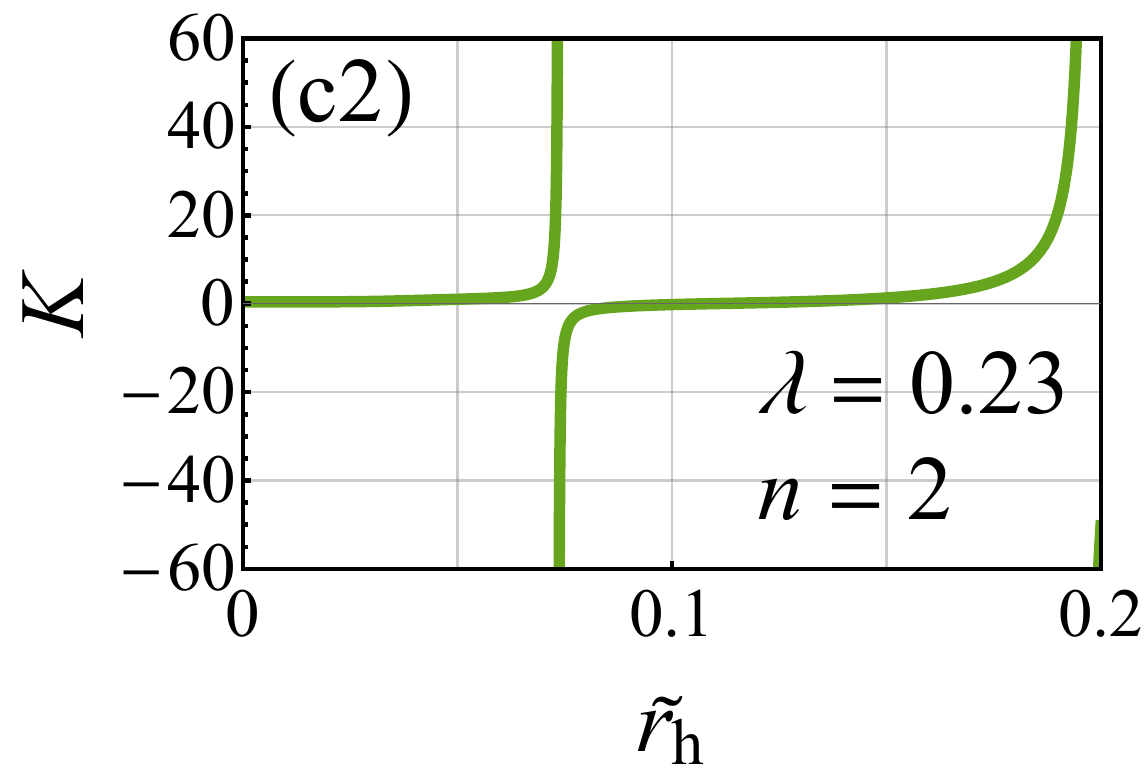}
  \end{minipage} }
  \hfill
\subfigure{\label{n33} \begin{minipage}[t]{0.3\textwidth}
    \centering
    \includegraphics[width=\textwidth]{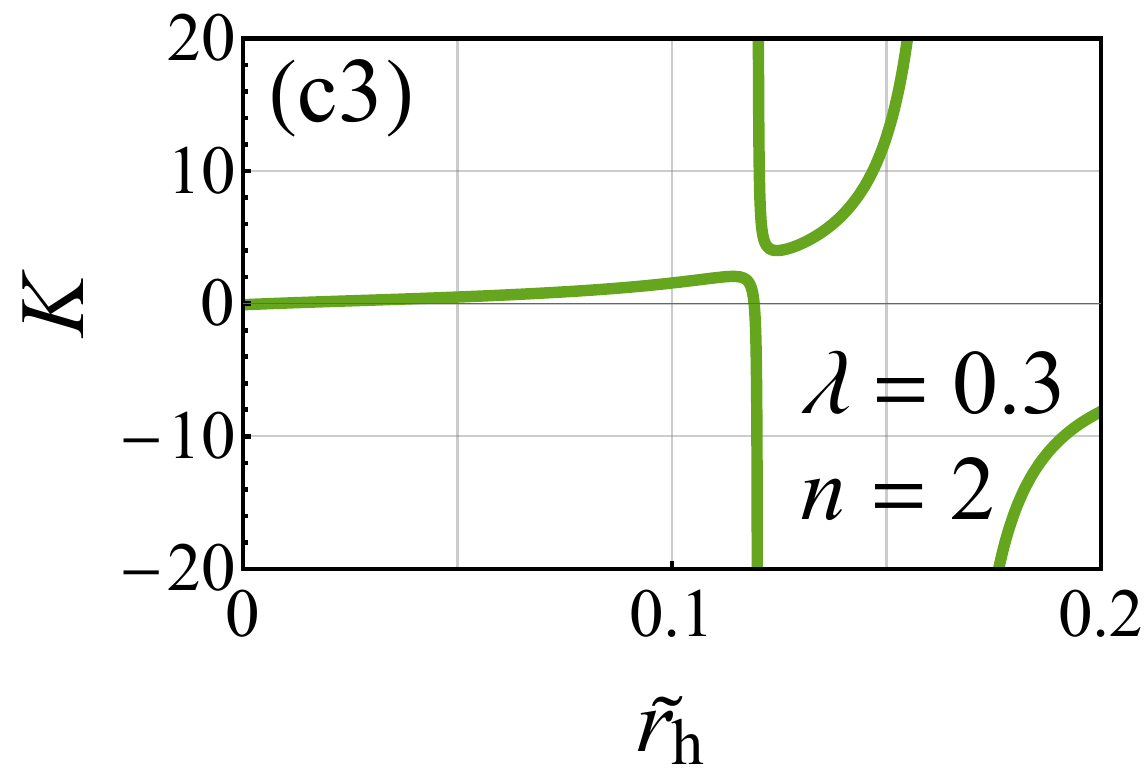}
  \end{minipage} }

 \caption{The QNMs frequencies slope $K = \frac{\mathrm{d}\omega_{\text{I}}/\mathrm{d}\tilde{r}_{\text{h}}}{\mathrm{d}\omega_{\text{R}}/\mathrm{d}\tilde{r}_{\text{h}}}$ as a function of the rescaled horizon radius $\tilde{r_h}$ under variation of the overtone number $n$ and the Euler-Heisenberg parameter $\lambda$ with the other parameters fixed on  $R_0=1$, $Q=0.5$, $f_{R_0}=0.1$ and $l=0$.}
  \label{fig:change n}
\end{figure}

It has been discussed that the QNMs spectrum of the $F(R)$-Euler-Heisenberg black holes exhibits structural changes as the black hole parameters vary. In this context, an increase in the overtone number may lead to different variations of the QNMs frequencies, especially the oscillatory part with the overtone number equaling to or exceeding the critical value, as illustrated in Fig. \ref{fig:omega to srh}. Consequently, the curves of the QNMs slope $K$ as a function of the rescaled horizon radius $\tilde{r}_{\text{h}}$ are significantly altered, as shown in Fig. \ref{fig:change n}. However, there still exist structure transitions when the black hole parameters are varied. 

To facilitate further analysis, we now define a new parameter $N_{\text{div}}$, which characterizes the number of the divergence points in the specific $K-\tilde{r}_{\text{h}}$ curve, 
\begin{equation*}
    N_{\text{div}}= \text{\textbf{the number of the divergence points in the specific $K-\tilde{r}_{\text{h}}$ curve}.}
\end{equation*}
In this way, we can investigate the influence of the overtone number $n$ by continuously varying the value of $\lambda$ and scanning the corresponding values of $N_{\text{div}}$.

\begin{figure}[htp!]
    \centering
    \includegraphics[width=0.55\linewidth]{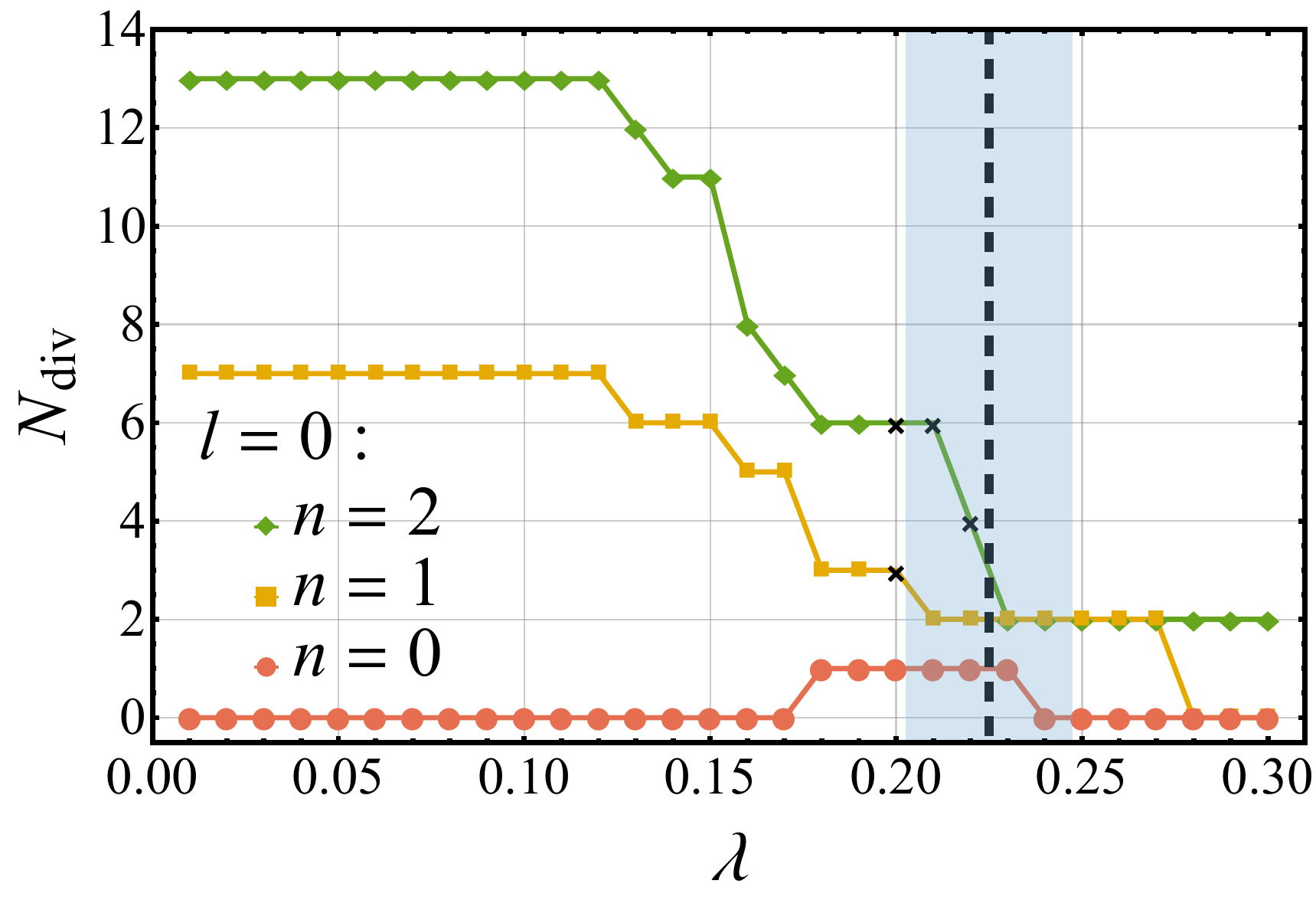}
    \caption{The number of the divergence points $N_{\text{div}}$ in the specific $K-\tilde{r}_{\text{h}}$ curve as a function of the Euler-Heisenberg parameter $\lambda$ under variation of the overtone number $n$ with the other parameters fixed on $R_0=1$, $Q=0.5$, $f_{R_0}=0.1$ and $l=0$. The cross marks in the figure indicate that the corresponding $K-\tilde{r}_{\text{h}}$ curves exhibit dramatic changes mentioned in Fig. \ref{o2}. The black dashed line indicates the correspond critical value of $\lambda$ taking $\lambda=0.225$ at which the heat capacity undergoes the structural transition under the same parameter settings, while the blue-shaded region represents the area where the absolute value of the relative error $\Delta_{\lambda}$ is less than $10\%$. }
    \label{fig:xl0}
\end{figure}

The results are described in Fig. \ref{fig:xl0}. It is shown that the oscillatory behavior in the QNMs frequencies caused by increasing the overtone number $n$ indeed lead to a significant increase in the number of the divergence points. However, it can also be observed that, near the critical parameters associated with the thermodynamic phase structure transitions in the black hole heat capacity discussed before, not only does the structure of the QNMs spectrum undergo a transformation (the corresponding positions have been marked with cross signs in the figure), but the characteristic structural change in the slope $K$ still persists. This is consistent with the results discussed above for the case of $n=0$, and it is reasonable to consider that the consistency observed between the structure transitions in heat capacity and in the slope of QNMs persists even when the overtone number $n$ is varied. 

\begin{figure}[htp!]
      \centering
  \subfigure{\label{mxl1} \begin{minipage}[t]{0.45\textwidth}
    \centering
    \includegraphics[width=\textwidth]{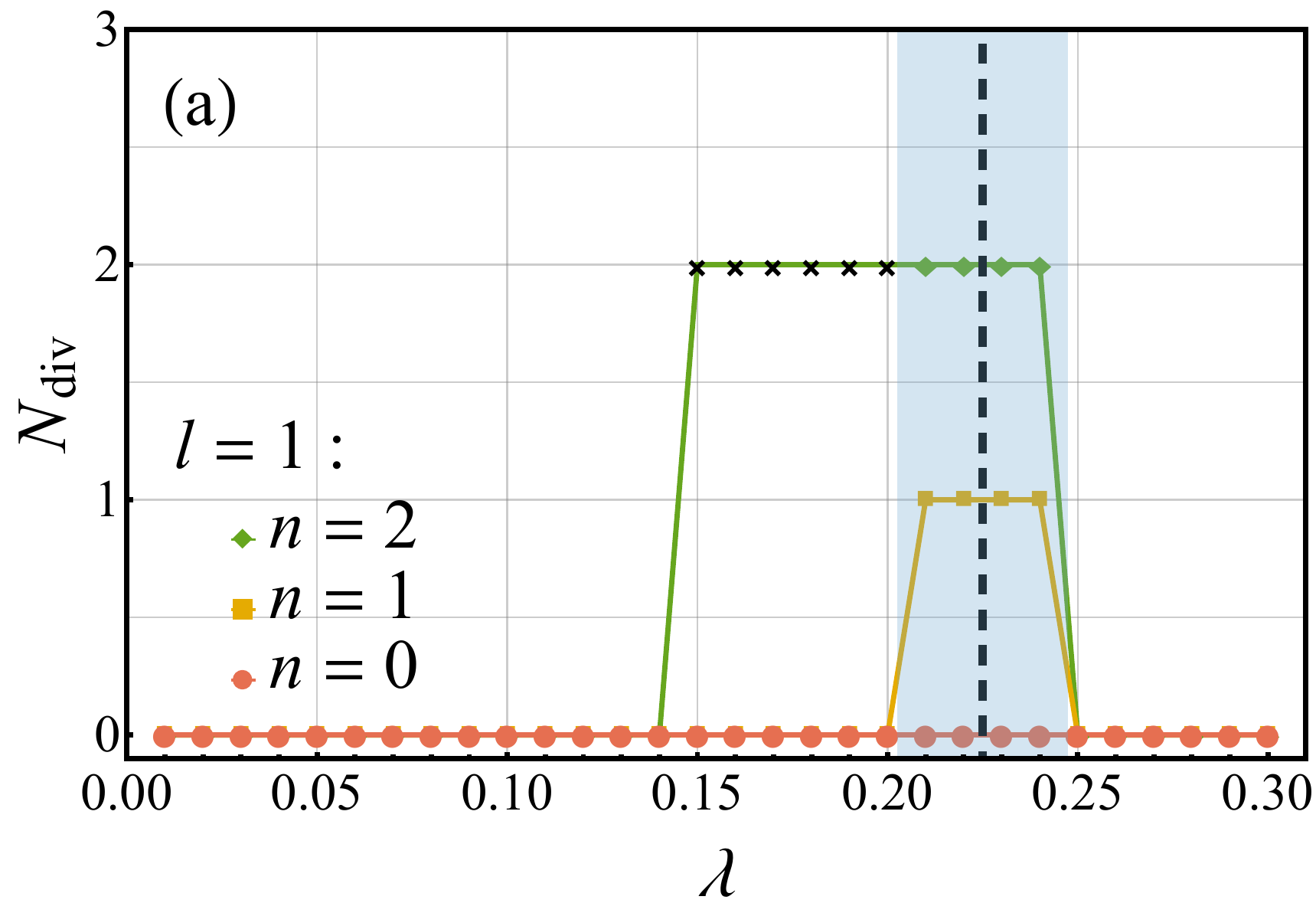}
  \end{minipage} }
\subfigure{\label{nxl2} \begin{minipage}[t]{0.45\textwidth}
    \centering
    \includegraphics[width=\textwidth]{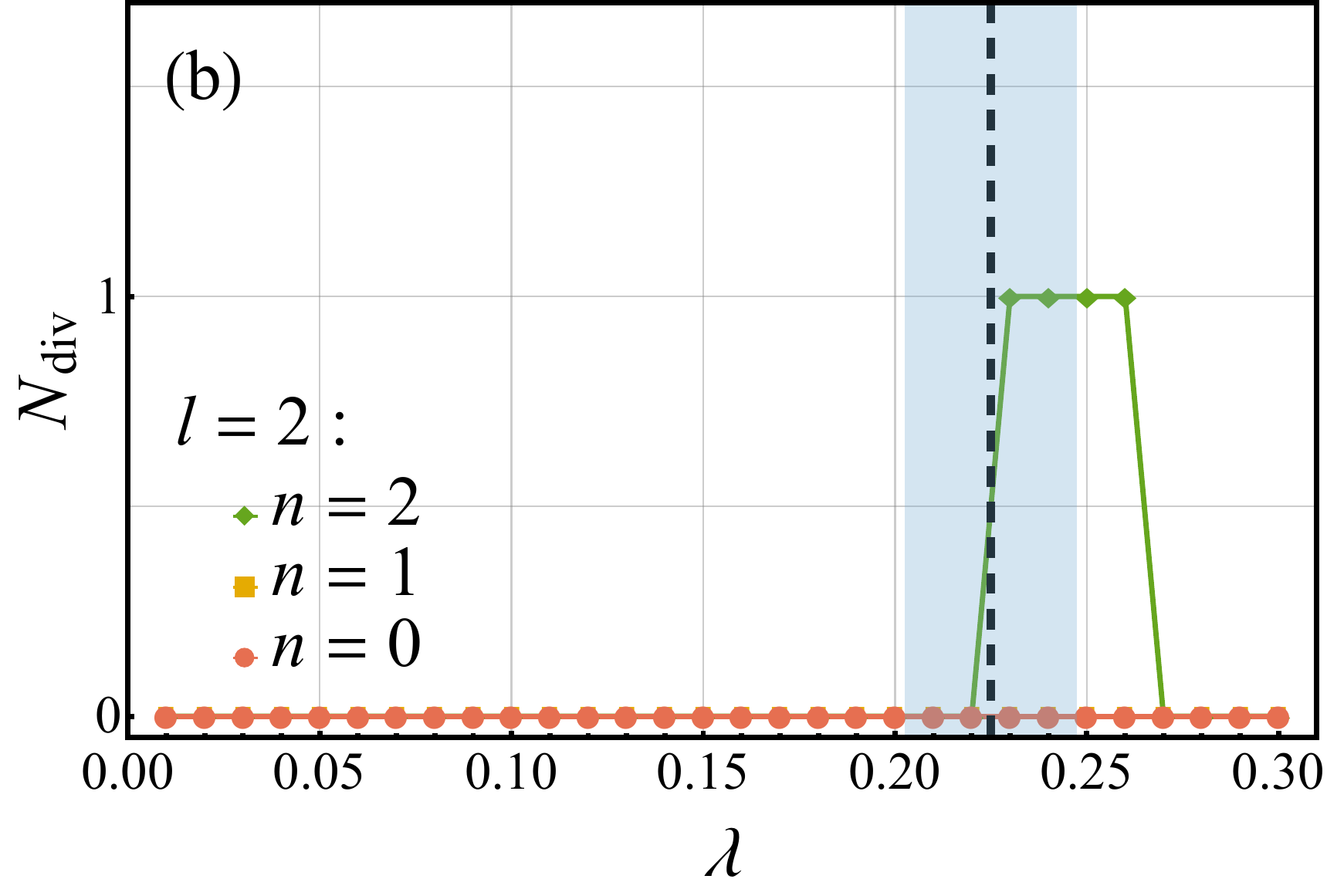}
  \end{minipage} }
  \hfill
 \caption{The number of the divergence points $N_{\text{div}}$ as a function of the Euler-Heisenberg parameter $\lambda$ under variation of the angular quantum number $l$ and the overtone number $n$ with the other parameters fixed on $R_0=1$, $Q=0.5$, $f_{R_0}=0.1$, in which the angular quantum number $l$ takes the value (a) $l=1$ and (b) $l=2$, respectively. The cross marks in the figure indicate that the corresponding $K-\tilde{r}_{\text{h}}$ curves exhibit discrete jump-like behaviors. The black dashed line indicates the correspond critical value of $\lambda$ at which the heat capacity undergoes the structural transition under the same parameter settings, while the blue-shaded region represents the area where the absolute value of the relative error $\Delta_{\lambda}$ is less than $10\%$.}
  \label{fig:xlc}
\end{figure}

Following the same methodology, we can further analyze the effects induced by varying the angular quantum number to $l\geq 1$. It is known that in certain cases the effects of increasing the angular quantum number may obscure some specific behaviors of the black hole system \cite{Song:2024kkx}, which is also the case in the QNMs. As shown in Fig. \ref{fig:xlc}, with the increase in $l$, the structural transitions characterized by the number of the divergence points $N_{\text{div}}$ disappear for small values of overtone number $n$. However, at this stage, if we try varying the overtone number, an interesting phenomenon will emerge. As $n$ increases, the previously obscured numerical variations in $N_{\text{div}}$ reemerge, and the reemerging variations are similar to the results observed in $l=0$. Moreover, for larger values of $l$, it can be found that the minimum value of $n$ at which this numerical variation reemerges is required to be larger. This indicates that there exists a competitive relationship between the angular quantum number $l$ and the overtone number $n$ in their effect on the behavior of QNMs, and the consistency between heat capacity and QNMs always persists intrinsically, although it may sometimes be obscured.

To uncover the underlying origin of this correspondence, we carried out a series of further investigations. In asymptotically AdS spacetimes, the Lyapunov exponent of the photon sphere has been verified to serve as a probe of thermodynamic phases \cite{Guo:2022kio,Yang:2023hci,Lyu:2023sih,Shukla:2024tkw,Lei:2024qpu,Chen:2025xqc,R:2025gok,Peng:2025htj,Awal:2025irl,Bezboruah:2025udi,Ali:2025ooh}. Besides, the effective potential base on the research between the QNMs of RNdS black holes and their Davies points\cite{Wei:2019jve,Tavlayan:2025asq} also serves the possible relationship between the black hole dynamics and thermodynamics. Specifically, we expand the calculations to the situation that we are considering, the $F(R)$-Euler-Heisenberg black hole with $R_0 > 0$, and the results are presented in detail in Appendix \ref{appendix3}. 
However, in asymptotically flat and de Sitter backgrounds, such correspondences are neither sufficiently clear nor widely validated. Clarifying the deeper physical mechanisms still warrant further and deeper investigation.

\ 

\section{Conclusions}\label{sec5}

In this work, we studied the properties of QNMs of the massless scalar field perturbation for the charged black holes in the $F(R)$-Euler-Heisenberg theory, which involves the setup of $F(R)$ gravity with nonlinear electrodynamic fields, and have also compared it with the black hole heat capacity at fixed charge. Our findings uncover a new correspondence between the structural behavior of QNMs and the structure of the heat capacity, suggesting the possible existence of a new connection between black hole thermodynamics and dynamics.

We first investigated the behavior of heat capacity and QNMs through varying parameter settings, and found that the critical point where the heat capacity changes from a single to double divergence nearly coincides with the point where the slope parameter $K$ of QNMs transitions from divergent to continuous and smooth behavior. Specifically, we choose the critical value of the Euler-Heisenberg parameter $\lambda$ obtained from the structural transition for comparison. By varying the values of the other three parameters $R_0$, $f_{R_0}$ and $Q$, we analyzed the numerical consistency of this correspondence, and the result shows that this consistency is widely present, including the case reducing to the standard Euler-Heisenberg-dS black hole limit (when $f_{R_0}=0$ and $R_0=4\Lambda$ with positive value). We calculated the relative error $\Delta_\lambda$ between the critical values of $\lambda$ at which the two different transitions occur, and further analyzed how the relative error is affected by individually varying the remaining parameters. We found that parameter space scaling induced by variations in different directions has a consistent effect on the relative error, indicating that the error is most likely of numerical origin.

We investigated the influence of changing the characteristics of the QNMs, the angular quantum number $l$ and the overtone number $n$, and found a competitive relationship between them. Specifically, the consistency between the QNMs and the thermodynamic phase structure transitions still exists, and at the same time, a larger angular quantum number $l$ tends to obscure the consistency, while a larger overtone number $n$ tends to produce the opposite effect. In other words, the competitive relationship between $n$ and $l$ leads to the result that for each specific value of the angular quantum number $l$, the consistency tends to be obscured when $n$ is small, but reemerges when $n$ exceeds a certain critical value.

For future work, investigating the QNMs of other types of perturbations would be an interesting direction, as it would not only test the validity of the aforementioned conclusions, but also allow us to examine the relationship between the angular quantum number and the overtone number in this phenomenon. It is also valuable to extend this work to include rotating black holes, which would be more realistic and could exhibit additional interesting features, and to examine whether our conclusions hold universally across different black hole backgrounds. Alternatively, one could investigate the relationship between black hole dynamics and thermodynamics in other contexts. It is also a natural and meaningful future direction to extend these results toward gravitational wave astronomy. Since QNMs can already be extracted from ringdown signals in gravitational wave observations and recent progress suggests that certain thermodynamic properties may already be verifiable through these astronomy signals \cite{Cardoso:2016rao,LIGOScientific:2016aoc,LIGOScientific:2025rsn,Gerosa:2020aiw,Wang:2023mst,Isi:2020tac,Nobili:2025ydt}, exploring whether the transitions measured from gravitational wave observations correlate with thermodynamic properties would potentially enrich black hole physics. 
Such future directions could offer potential insights into the connection between black hole thermodynamics and dynamics, possibly leading to new tests of fundamental physics. 

\section*{Acknowledgement} We would like to thank Bum-Hoon Lee, Qi-Yuan Pan, Wen-Bin Li, Surojit Dalui and Shuta Ishgaki for helpful discussions. The work of XHG is supported in part by NSFC, China (Grant No. 12275166 and No. 12311540141). YQL is partly supported by NSFC, China (Grant No.12405072) and China Postdoctoral Science Foundation (Grant No. 2024M761914).



\appendix
\section{$F(R)$-Euler-Heisenberg Black Hole Geometry}\label{appendix1}

\begin{figure}[ht!]
\centering

\subfigure{\label{bf1}
\begin{minipage}[t]{0.31\linewidth}
\centering
\includegraphics[width=1\textwidth]{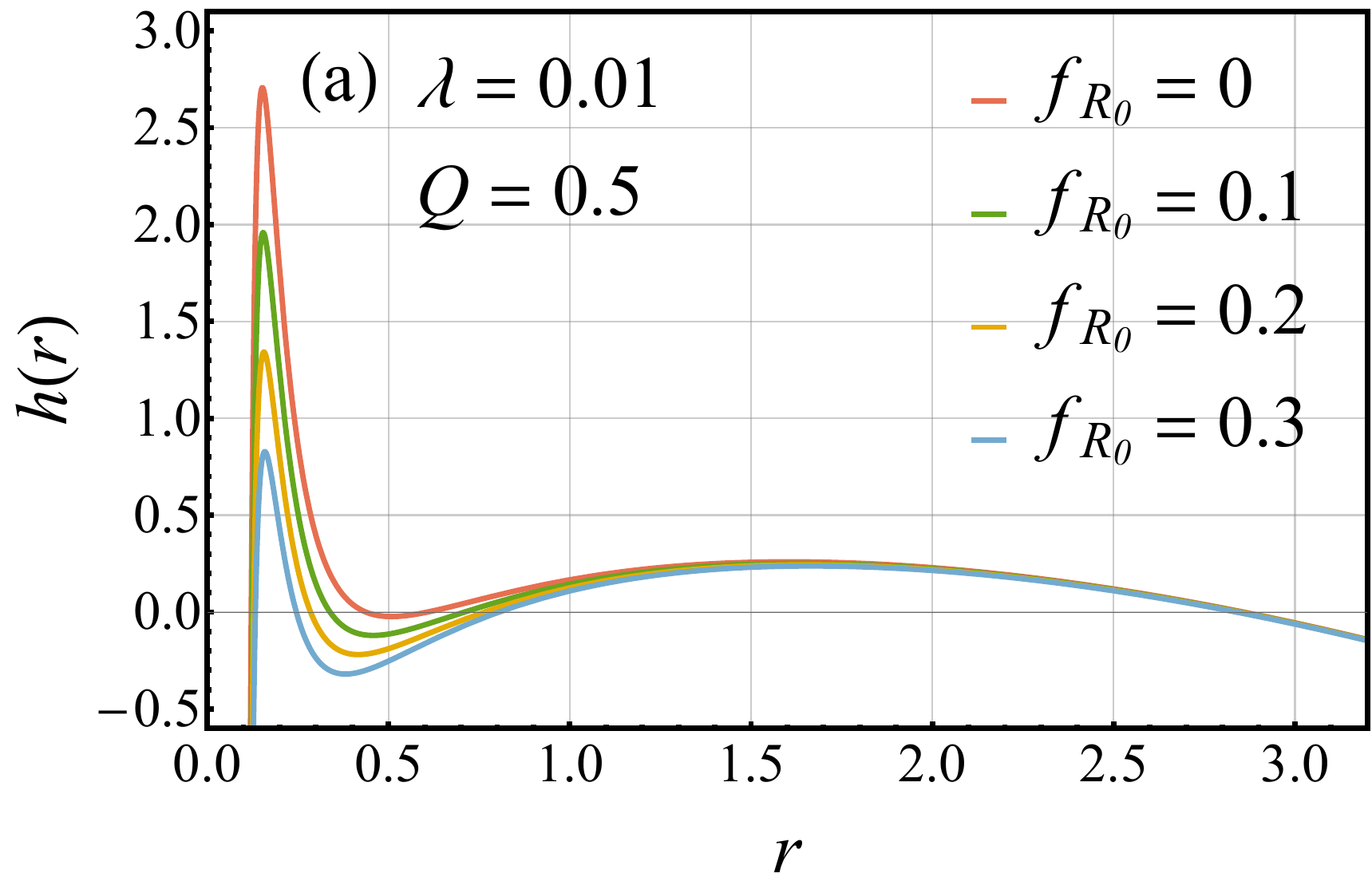}
\end{minipage}
}
\subfigure{\label{bf2}
\begin{minipage}[t]{0.31\linewidth}
\centering
\includegraphics[width=1\textwidth]{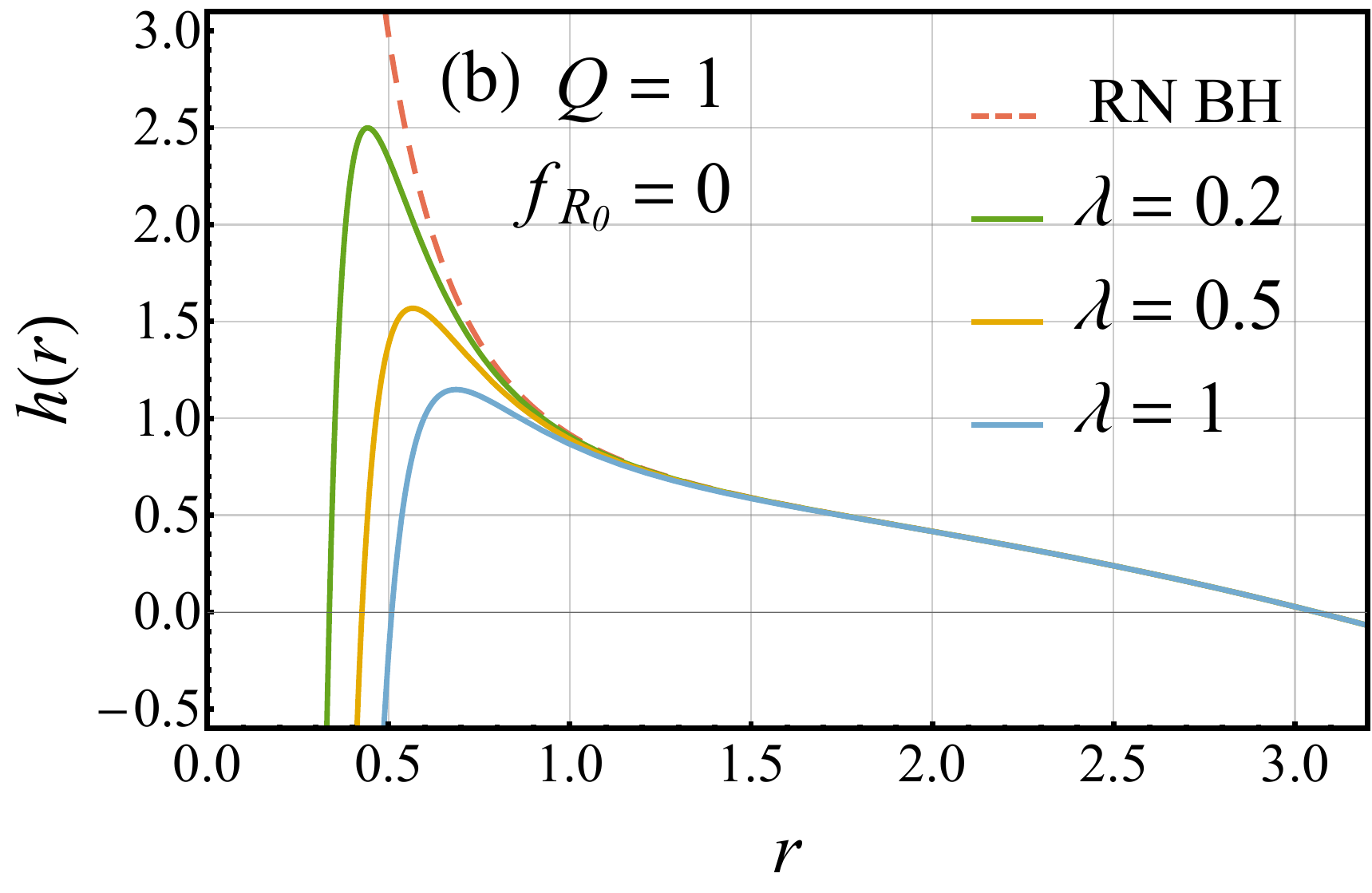}
\end{minipage}
}
\subfigure{\label{bf3}
\begin{minipage}[t]{0.31\linewidth}
\centering
\includegraphics[width=1\textwidth]{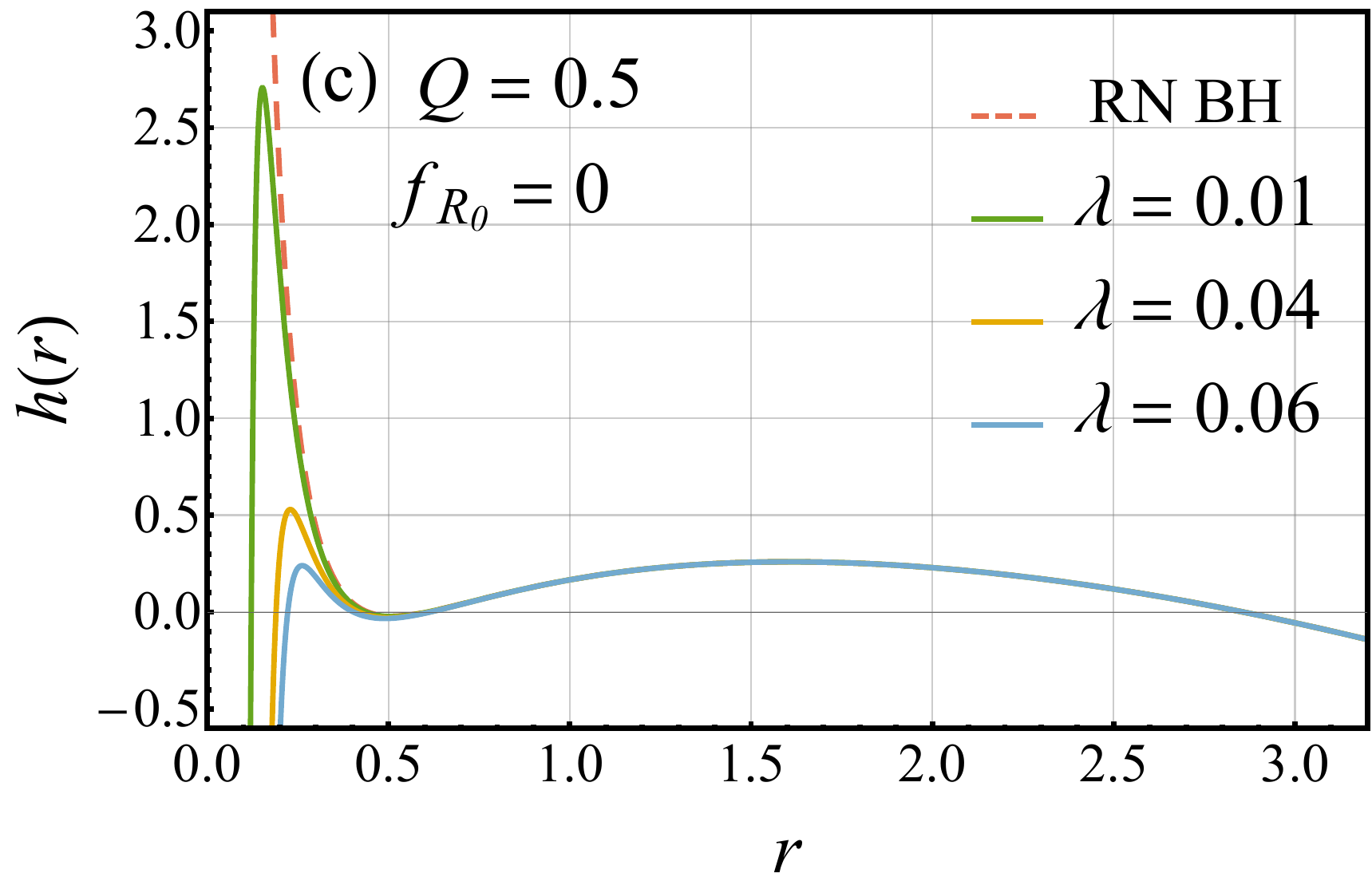}
\end{minipage}
}

\caption{The blackening factor $h(r)$ of the $F(R)$-Euler-Heisenberg black holes varying $f_{R_0}$ and $\lambda$. The dashed curves are for RN-dS black holes ($\lambda=0$ and $f_{R_0}=0$). Figure (a) are for different values of $f_{R_0}$. Figures (b) and (c) are for different values of the Euler-Heisenberg parameter $\lambda$. These are for fixed value of $m_0=1$, $R_0=1$.}
\label{fig:blackening}
\end{figure}

The variation of the blackening factor $h(r)$ in the Eq.(\ref{eq:metric func}) of the $F(R)$-Euler–Heisenberg black hole under different parameter changes is illustrated in Fig. \ref{fig:blackening}. As discussed previously, there may be two, three, or even four horizons, with one of them being the cosmological horizon. For $f_{R_0}$, as shown in Fig. \ref{bf1}, the variation mainly influences the region inside the black hole event horizon. As $f_{R_0}$ gradually increases, the maximum value of $h(r)$ inside the black hole event horizon decreases, and the two inner horizons, $r_{\text{i}}$ and $r_{\text{i}^{\prime}}$, of the black hole gradually approach each other. Without considering the effect of $f_{R_0}$, i.e., taking $f_{R_0}=0$ as shown in Fig. \ref{bf2} and \ref{bf3}, the variation of the Euler-Heisenberg parameter $\lambda$ significantly influences the behavior of the blackening factor $h(r)$ at very small values of $r$. As $\lambda$ increases, $h(r)$ decreases rapidly. However, outside this regime, the black hole structure is almost identical to that of an RN-dS black hole with the same charge and mass (the dashed red curves).

Such results are expected, since the sign of the nonlinear part of the electromagnetic term in Eq.(\ref{eq:metric func}) is opposite to that of the linear part, and the power of $r$ in it is $-6$. Hence, the nonlinear term directly affects the behavior and divergence direction of $h(r)$ in the vary small-$r$ regime, while small variations of $f_{R_0}$ hardly affect the structure for which it only appears as a coefficient of the electromagnetic term in the form $1/(1+f_{R_0})$. The blackening factor $h(r)$ can thus be divided into three regimes: the region where the nonlinear electromagnetic term dominates at very small $r$, the region where the linear electromagnetic term dominates at moderately small $r$, and the region where the mass term and the cosmological parameter dominate at large $r$. The dominant behavior of the solution is still governed by linear theory.

\section{Chebyshev Pseudo-Spectral Method}\label{appendix2}

The expressions for the coefficient of Eq.(\ref{eq:final eq}) are as follows,

\begin{dmath}
    A_0(v)=- u_{\text{c}}^4 v^2 \left( u_{\text{c}} + u_{\text{b}} v - u_{\text{c}} v \right)^3 h\left( u_{\text{c}} + u_{\text{b}} v - u_{\text{c}} v \right) \left( h^{\prime}(u_{\text{c}}) \right)^2,
\end{dmath}

\begin{dmath}
    B_0(v)=u_{\text{c}}^4 v \left( u_{\text{c}} (-1 + v) - u_{\text{b}} v \right)^2 \left( u_{\text{c}} + u_{\text{b}} v - u_{\text{c}} v \right) \left( h^{\prime}(u_{\text{c}}) \right)^2 \left[ 2 h\left( u_{\text{c}} + u_{\text{b}} v - u_{\text{c}} v \right) + \left( -u_{\text{b}} + u_{\text{c}} \right) v\, h^{\prime}\left( u_{\text{c}} + u_{\text{b}} v - u_{\text{c}} v \right) \right],
\end{dmath}

\begin{dmath}
    B_1(v)= -2 u_{\text{c}}^2 v \left( u_{\text{c}} (-1 + v) - u_{\text{b}} v \right) h^{\prime}(u_{\text{c}}) \left[ 2 \left( u_{\text{c}} (-1 + v) - u_{\text{b}} v \right)^2 h\left( u_{\text{c}} + u_{\text{b}} v - u_{\text{c}} v \right) + u_{\text{c}}^2 (-u_{\text{b}} + u_{\text{c}}) v\, h^{\prime}(u_{\text{c}}) \right],
\end{dmath}

\begin{dmath}
    C_0(v)=u_{\text{c}}^4 \left( u_{\text{c}} + u_{\text{b}} v - u_{\text{c}} v \right) h^{\prime}(u_{\text{c}})^2 \left[ -2 \left( u_{\text{c}} (-1 + v) - u_{\text{b}} v \right)^2 h\left( u_{\text{c}} + u_{\text{b}} v - u_{\text{c}} v \right) + (u_{\text{b}} - u_{\text{c}}) v \left( l (1 + l)(u_{\text{b}} - u_{\text{c}}) v + \left( u_{\text{c}} + u_{\text{b}} v - u_{\text{c}} v \right)^2 h^{\prime}\left( u_{\text{c}} + u_{\text{b}} v - u_{\text{c}} v \right) \right) 
\right],
\end{dmath}

\begin{dmath}
    C_1(v)=2 u_{\text{c}}^2 h^{\prime}(u_{\text{c}}) \left[3 \left( u_{\text{c}} (-1 + v) - u_{\text{b}} v \right)^3 h\left( u_{\text{c}} + u_{\text{b}} v - u_{\text{c}} v \right) + (u_{\text{b}} - u_{\text{c}}) v \left( u_{\text{c}}^2 (u_{\text{c}} + 2 u_{\text{b}} v - 2 u_{\text{c}} v) h^{\prime}(u_{\text{c}}) + (u_{\text{c}} + u_{\text{b}} v - u_{\text{c}} v)^3 h^{\prime}\left( u_{\text{c}} + u_{\text{b}} v - u_{\text{c}} v \right) \right)
\right],
\end{dmath}

\begin{dmath}
    C_2(v)=-4 \bigl( u_{\text{c}} (-1 + v) - u_{\text{b}} v \bigr) \left[ \bigl( u_{\text{c}} (-1 + v) - u_{\text{b}} v \bigr)^2 \, h\bigl( u_{\text{c}} + u_{\text{b}} v - u_{\text{c}} v \bigr) + u_{\text{c}}^2 (-u_{\text{b}} + u_{\text{c}}) v \, h^{\prime}(u_{\text{c}}) \right],
\end{dmath}
where $u_{\text{b}}=1/r_{\text{h}}$ and $u_{\text{c}}=1/r_{\text{c}}$ are the black hole event horizon and the cosmological horizon, respectively, $v=(u - u_{\text{c}})/(u_{\text{b}} - u_{\text{c}})$ is a transformed coordinate and $h(u)$ is the blackening factor expressed as a function of $u$, while the prime $^{\prime}$ stands for $\frac{\mathrm{d}}{\mathrm{d}u}$.

In this expression, $A$, $B$ and $C$ correspond to the coefficients of the derivatives of the second, first and zero order of $\varphi$ with respect to $v$, and the subscripts 0, 1, and 2 indicate their dependence on $\omega^0$, $\mathrm{i}\omega^1$ and $\omega^2$, respectively.
Expanding $v$ on the Chebyshev grid $\hat{v}=\{v_j=\cos\left( \frac{j \pi}{N} \right),\ j = 0, 1, 2, \dots, N\}$ with $N$ the freely chosen total number of nodes, and denoting the corresponding $\varphi$, first-order derivatives ${\mathrm{d}\varphi}/{\mathrm{d}v}$ and second-order derivatives ${\mathrm{d}^2\varphi}/{\mathrm{d}v^2}$ in terms of Chebyshev spectral modes $\hat{D}_0$, $\hat{D}_1$ and $\hat{D}_2$, respectively, the differential equation Eq.(\ref{eq:final eq}) can thus be transformed into a matrix equation,
\begin{equation}
\Big(\hat{A}_0\hat{D}_2+\hat{B}_0\hat{D}_1+\hat{C}_0\hat{D}_0\Big)+\mathrm{i}\omega\Big(\hat{B}_1\hat{D}_1+\hat{C}_1\hat{D}_0\Big)+\omega^2\hat{C}_2\hat{D}_0=0,
\end{equation}
in which $\hat{A}_0$, $\hat{B}_1$, $\hat{B}_0$, $\hat{C}_2$, $\hat{C}_1$, and $\hat{C}_0$ are the correspondent matrices of $A_0(\hat{v})$, $B_1(\hat{v})$, $B_2(\hat{v})$, $C_2(\hat{v})$, $C_1(\hat{v})$ and $C_0(\hat{v})$, respectively. The problem of solving QNMs frequencies is now transform into a matrix eigenvalue problem, which can be easily solved numerically.

\section{Lyapunov Exponents and Effective Potential}\label{appendix3}
In this section, we present the Lyapunov exponent of the photon sphere and the effective potential of the QNMs K-G equation, attempting to verify the relationship between QNMs and thermodynamics from them. Although these quantities do not exhibit a direct connection with the transitions of the thermodynamic phase structure, they still serve as valuable references and merit further investigation in future work.

We start from the line element of a static, spherically symmetric black hole, $ds^{2} = -h(r)\, dt^{2} + \frac{dr^{2}}{h(r)} + r^{2} d\Omega^{2}$.
The corresponding Lagrangian for a test particle on the equatorial plane ($\theta=\pi/2$) is
\begin{equation}
    \mathcal{L} = \frac{1}{2} g_{\mu\nu} \dot{x}^{\mu} \dot{x}^{\nu}
    = \frac{1}{2}\left[-h(r)\dot{t}^{2} + \frac{\dot{r}^{2}}{h(r)} + r^{2}\dot{\phi}^{2}\right].
\end{equation}
The associated conserved quantities are $E = h(r)\dot{t}$, $L = r^{2}\dot{\phi}.$ The radial equation then becomes
\begin{equation}
    \dot{r}^{2} + \mathcal{V}_{\text{eff}}(r) = 0, \ \ \    \mathcal{V}_{\text{eff}}(r) = -E^{2} + h(r)\left(\frac{L^{2}}{r^{2}} - \eta \right),
\end{equation}
where $\eta=0$ corresponds to null geodesics and $\eta=1$ corresponds to timelike geodesics.

Setting $\eta = 0$, the effective potential for photons simplifies to $\mathcal{V}_{\text{eff}}(r) = -E^{2} + \frac{L^{2}}{r^{2}} h(r)$. The photon circular orbit is determined by
\begin{equation}
    \mathcal{V}_{\text{eff}}(r)=0, \qquad 
    \frac{d\mathcal{V}_{\text{eff}}}{dr} = 0.
\end{equation}
From $\mathcal{V}_{\text{eff}}=0$, we obtain the constraint $\frac{E^{2}}{L^{2}}= \frac{h(r)}{r^{2}}$.
The condition for extremizing $\mathcal{V}_{\text{eff}}$ yields the photon-sphere equation $-\frac{2 h(r)}{r} + h'(r) = 0$, and the root gives the photon sphere radius $r = r_{\text{ph}}$.

For null geodesics, the Lyapunov exponent describing the instability of the photon orbit is
\begin{equation}
    \lambda_L = \sqrt{-\frac{\mathcal{V}_{\text{eff}}''(r_{\text{ph}})}{2\,\dot{t}^{\,2}}}.
\end{equation}
This formula is used throughout the main text when evaluating the dynamical instability of photon circular orbits and comparing with thermodynamic phase structures.

\begin{figure}[H]
    \centering
    \includegraphics[width=.5\textwidth]{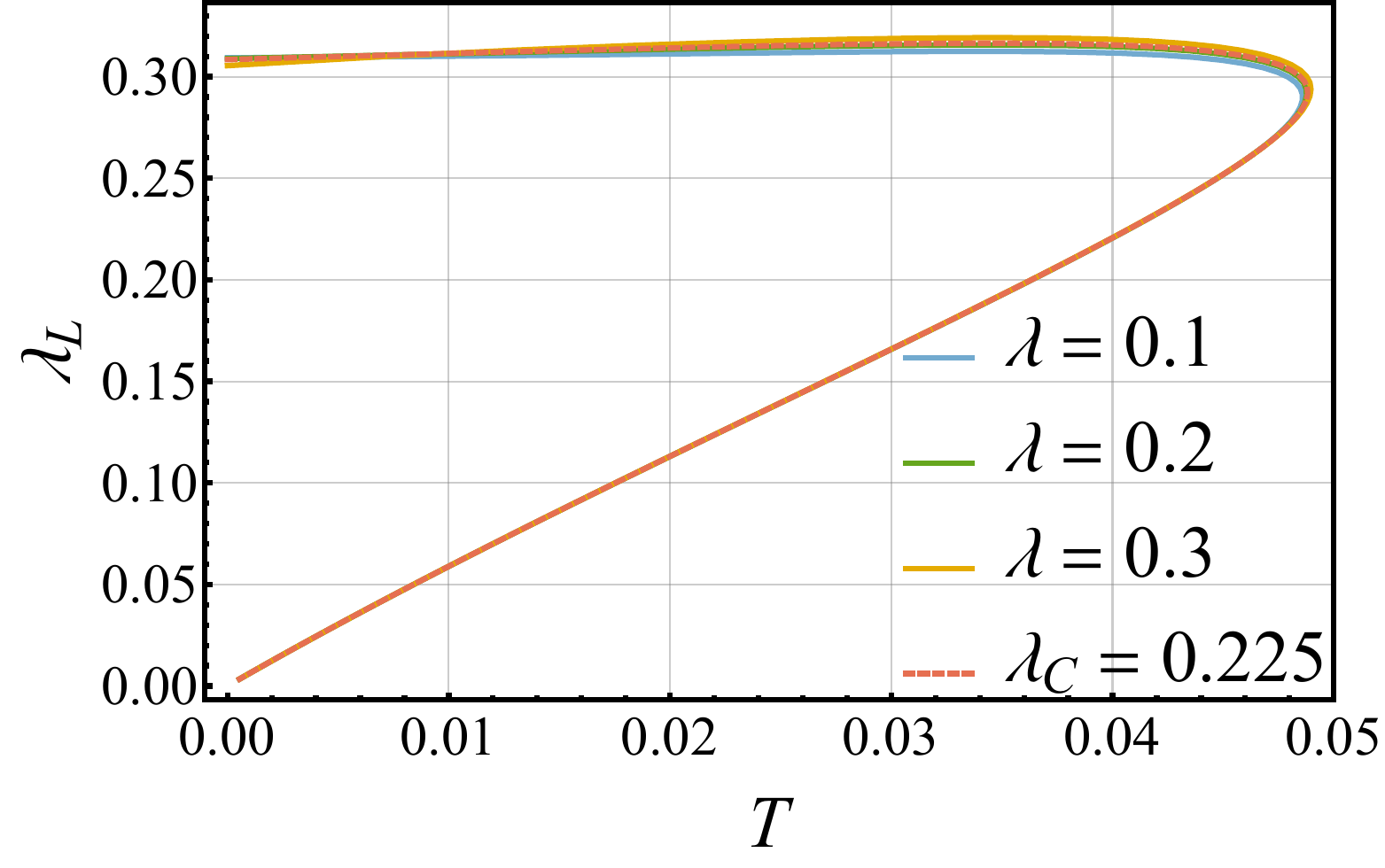}
    \caption{Lyapunov exponents versus black hole temperature under different values of Euler-Heisenberg parameter $\lambda$. The values of $\lambda$ are $0.1$, $0.2$ and $0.3$, respectively. The red dashed line stands for $\lambda_C=0.225$, the transition value of $\lambda$ between different capacity phase structures. The other parameters are fixed on $R_0=1$, $Q=0.5$ and $f_{R_0}=0.1$. }
     \label{fig.Lyapunov}
\end{figure}

The specific computational results are shown in Fig. \ref{fig.Lyapunov}. We find that, unlike some previous results in AdS backgrounds, the Lyapunov exponent does not signal thermodynamic phase transitions in the system and does not exhibit any apparent connection with changes in the thermodynamic phase structure.

\begin{figure}[!ht]
\centering

\subfigure{\label{3d1}
\begin{minipage}[t]{0.3\linewidth}
\centering
\includegraphics[width=1\textwidth]{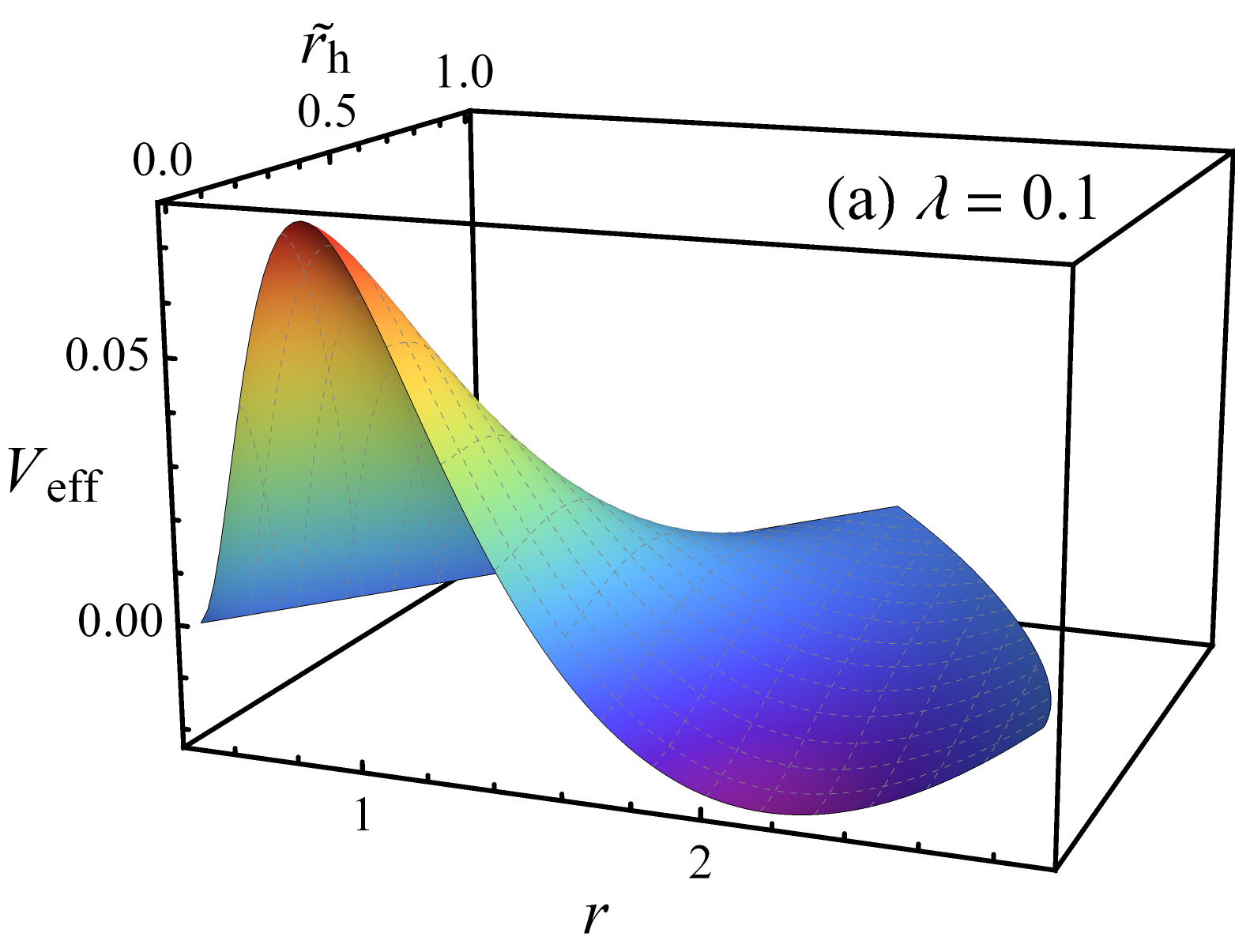}
\end{minipage}
}
\subfigure{\label{3d2}
\begin{minipage}[t]{0.3\linewidth}
\centering
\includegraphics[width=1\textwidth]{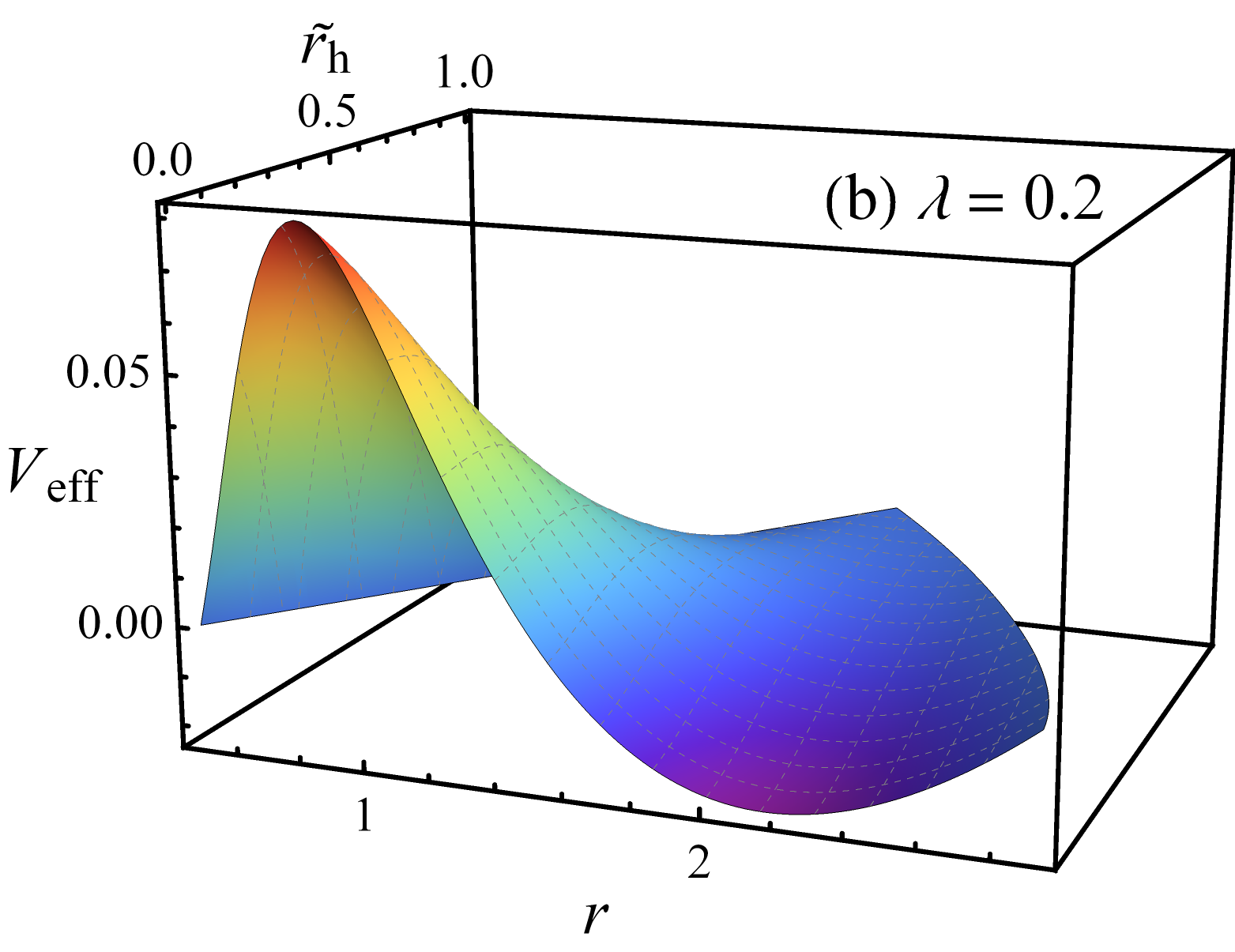}
\end{minipage}
}
\subfigure{\label{3d3}
\begin{minipage}[t]{0.3\linewidth}
\centering
\includegraphics[width=1\textwidth]{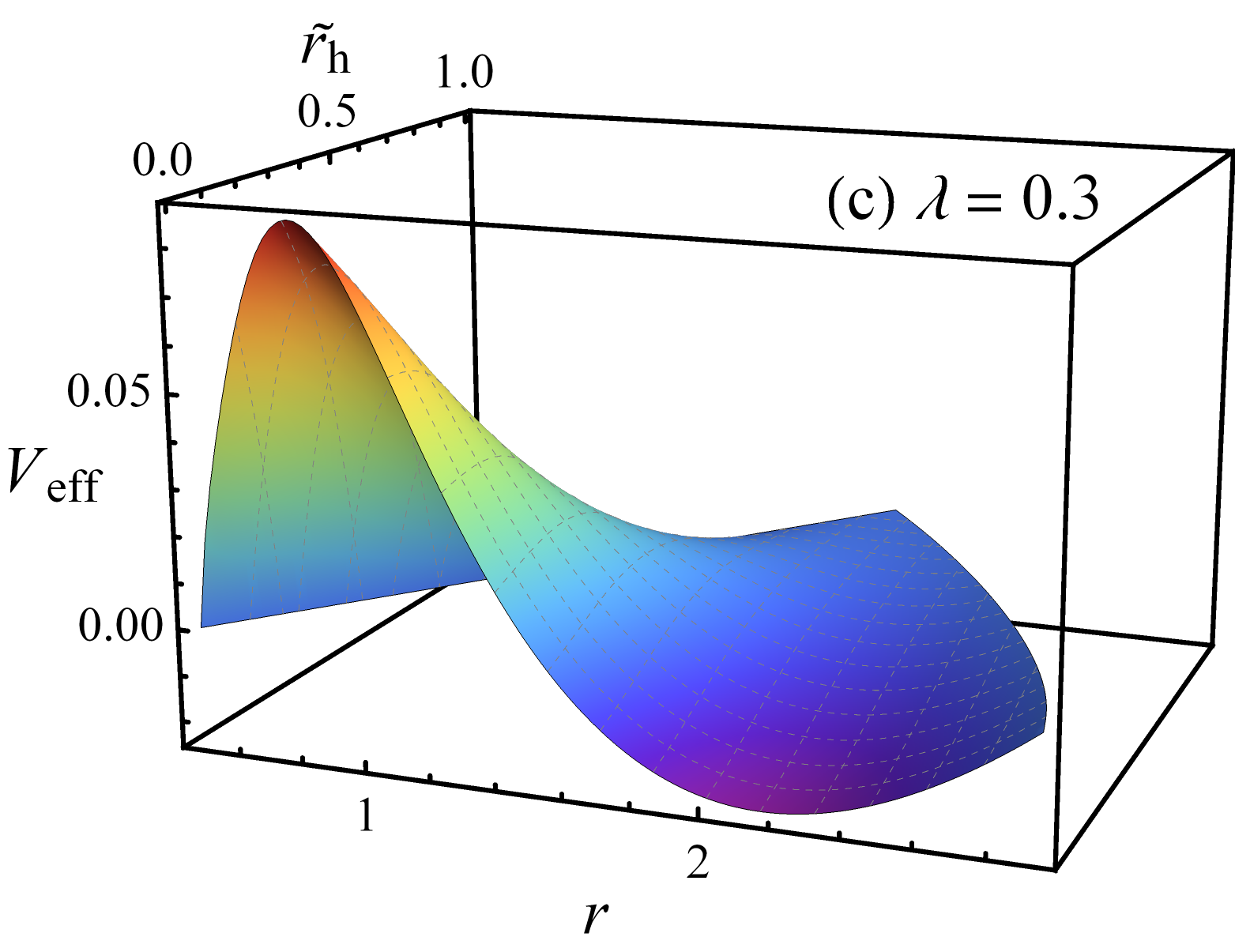}
\end{minipage}
}

\caption{Color map indicates the magnitude of effective potential $V_{\text{eff}}(r,\tilde{r}_{\text{h}})$ with the angular quantum number $l=0$, where $\tilde{r}_{\text{h}}$ is the rescaled horizon radius defined by $\tilde{r}_{\text{h}}=({r_{\text{Nariai}}-r_{\text{h}}})/({r_{\text{Nariai}}-r_{\text{extreme}}})$.}
\label{fig:3dveff}
\end{figure}

For the QNMs effective potential, the function is 
\begin{equation}
    V_{\text{eff}}(r) = h(r) \left( \frac{\ell(\ell+1)}{r^2} + \frac{h^{\prime}(r)}{r} \right),
\end{equation}
which is from Eq.(\ref{eq: QNMs like-Sch}). Fixing the angular quantum number $l=0$, we plotted the effective potential with different values of the Euler-Heisenberg parameter $\lambda$, as shown in Fig. \ref{fig:3dveff}. The results show no direct significant change. In particular, the three different values of $\lambda$ correspond to the the three distinct heat capacity structures shown in Fig. \ref{fig:CQlines}. However, these differences are not visible in the plotted maps in Fig. \ref{fig:3dveff}, as the behaviors appear nearly indistinguishable at the graphical level.

We believe that the correspondence observed in the manuscript between the transition of QNM behavior and the change in the thermodynamic phase structure may be caused by additional underlying physical mechanisms that have not yet been fully understood.

\end{document}